\documentclass[conference]{IEEEtran}

\usepackage{booktabs}
\usepackage{cite}
\usepackage{amsmath,amssymb,amsfonts}
\usepackage{textcomp}
\usepackage{xcolor}

\usepackage{graphicx}
\usepackage{algorithm}
\usepackage[noend]{algpseudocode}
\usepackage[caption=false]{subfig}
\usepackage{comment}
\usepackage{enumitem}
\usepackage{nameref}
\usepackage{multirow}
\usepackage{array}
\usepackage{hyperref}
\usepackage{cleveref}

\begin{document}

\title{Bit-Accurate Modeling of GPU Matrix Multiply-Accumulate Units: Demystifying Numerical Discrepancy and Accuracy}

\author{\IEEEauthorblockN{Peichen Xie}
\IEEEauthorblockA{Microsoft Research}
\and
\IEEEauthorblockN{Shuotao Xu}
\IEEEauthorblockA{Microsoft Research}
\and
\IEEEauthorblockN{Yang Wang}
\IEEEauthorblockA{Microsoft Research}
\and
\IEEEauthorblockN{Fan Yang}
\IEEEauthorblockA{Microsoft Research}
\and
\IEEEauthorblockN{Mao Yang}
\IEEEauthorblockA{Microsoft Research}
}

\maketitle
\thispagestyle{plain}
\pagestyle{plain}

\begin{abstract}
Modern AI accelerators rely on matrix multiply-accumulate units (MMAUs), such as NVIDIA Tensor Cores and AMD Matrix Cores, to accelerate deep neural network workloads. MMAUs expose only instruction-level or API-level interfaces of matrix multiply-accumulate (MMA) operations, while leaving internal floating-point arithmetic behavior undocumented. Consequently, MMAUs across vendors and architectural generations often produce numerical discrepancies for identical inputs, and sometimes exhibit reduced numerical accuracy that can cause training instability. Diagnosing and understanding the root causes of these effects is challenging without white-box models of their arithmetic behavior. This paper proposes closed-loop feature probing (CLFP), a generic and systematic framework for constructing bit-accurate arithmetic behavior models of MMA operations. Based on this framework, we analyze all MMA instructions on ten GPU architectures spanning NVIDIA Volta through RTX Blackwell and AMD CDNA1 through CDNA3, and derive the first bit-accurate arithmetic models for these MMAUs. Our models explain previously observed cross-platform numerical discrepancies and accuracy issues, enable white-box numerical error analysis, reveal four types of precision bottlenecks and one type of numerical asymmetry, and inform software workarounds as well as design suggestions for future MMAUs. This work is open-source at \url{https://github.com/microsoft/MMA-Sim}.
\end{abstract}

\section{Introduction}

Deep neural networks (DNNs) are built on massive numbers of linear algebra operations. 
To satisfy the rapidly growing computational demands of modern DNN workloads such as large language models (LLMs), recent AI accelerators have introduced specialized hardware units such as NVIDIA Tensor Cores \cite{markidis_nvidia_2018} and AMD Matrix Cores \cite{schieffer_rise_2024} to accelerate matrix multiplications. 
These \textit{matrix multiply-accumulate units} (MMAUs) now dominate both training and inference workloads because of their extremely high throughput, efficiency, and support for mixed-precision arithmetic. 
As DNN workloads scale in model size and deployment scope, the numerical behavior of MMAUs becomes increasingly critical since small discrepancies can accumulate across a large number of operations and affect reproducibility and training stability (\S\ref{sec:background}).

Despite nominally performing the same mathematical operation, MMAUs from different AI accelerators are not numerically equivalent. 
Even with bit-identical inputs, different MMAU implementations can produce different floating-point results. 
Such discrepancies are pervasive across vendors \cite{li_discovery_2024} and also occur across architectural generations from the same vendor \cite{fasi_numerical_2021}. 
These hardware-level numerical discrepancies could propagate through the software stack and manifest as inconsistencies in end-to-end DNN workloads, resulting in poor reproducibility across different AI accelerators.

Compounding the problem, the underlying \textit{arithmetic behavior} of MMAUs, i.e., the hardware-specific floating-point computation mechanism for matrix multiply-accumulate (MMA) operations, is poorly documented and effectively a black box, leaving developers without a clear understanding of how numerical results are produced. 
As a result, the effect of arithmetic behavior on numerical accuracy is often discovered only after it surfaces in large-scale systems.
Recent incidents illustrate this issue. 
Researchers and developers reported that certain MMA instructions on NVIDIA Hopper GPUs and AMD CDNA2 GPUs degraded training accuracy and stability due to insufficient summation precision or subnormal flushing \cite{deepseek-ai_deepseek-v3_2024,zhao_insights_2025,pytorch_developers_pytorch_2025,zhang_sageattention2_2025}.
In both cases, the root causes lie in the obscure arithmetic behavior of MMAUs.

These observations reveal a fundamental gap: the lack of a white-box understanding of MMAU arithmetic behavior. 
Without such a foundation, numerical analysis remains \textit{empirical} and \textit{reactive}, making it difficult to predict accuracy, diagnose issues, or design principled mitigations. 
A systematic white-box model of MMAU arithmetic behavior is therefore essential for \textit{rigorous} and \textit{proactive} reasoning about the numerical behavior of modern AI systems.

However, obtaining a white-box model for MMAUs is inherently difficult. 
MMAUs expose only high-level MMA interfaces at the API level or at the instruction level, while the arithmetic features that determine numerical results remain hidden inside proprietary hardware. 
These features include computational order, fusion granularity, internal precision, rounding behavior, and many other subtle design choices.
Because the relevant feature set is neither explicitly documented nor straightforward to enumerate exhaustively, the space of possible arithmetic behavior is effectively enormous, making exhaustive black-box exploration challenging.
For example, prior feature-probing studies infer isolated or partial arithmetic features for a few architectures, but none can provide a model that covers the complete feature set and is verified end-to-end for bit accuracy (\S\ref{sec:background}).

In this paper, we tackle this challenge with a generic, systematic, and principled approach for constructing white-box models of MMA arithmetic behavior through black-box testing. 
In particular, we propose \textit{closed-loop feature probing} (CLFP), a framework that integrates arithmetic feature probing with iterative refinement (\S\ref{sec:method}). 
Given an MMA interface, we carefully design edge-case inputs to probe arithmetic features. 
Based on the probed features, we build an executable arithmetic behavior model and then iteratively refine it through a verify-revise loop until the model is verified to reproduce the outputs of the MMA interface bit-for-bit.
This closed-loop methodology enables complete and accurate modeling of black-box MMAU arithmetic behavior. 

In this work, we apply CLFP to \textit{instruction-level interfaces} of GPU MMAUs.
Using the CLFP framework, we systematically analyze \textit{all floating-point MMA instructions on ten GPU architectures}, spanning NVIDIA's architectures from Volta to RTX Blackwell and AMD's architectures from CDNA1 to CDNA3, and construct white-box models of their arithmetic behavior (\S\ref{sec:model}).
Our models accurately reveal the complete arithmetic behavior of MMAUs for the first time, showing that their MMA operations are composed of different types of elementary floating-point operations with different precisions, rounding modes, computational orders, and fusion granularities.

These white-box models explain previously observed numerical discrepancies across accelerators and expose the design choices that can affect numerical accuracy. For example, the models explain how the MMAUs produce six different output values ($0.0$, $-0.375$, $-0.5$, $-0.75$, $-0.875$, and $-1.0$) for the same input (\S\ref{sec:discrepancy}). Our white-box models also enable us to quantitatively analyze the sources of numerical errors, revealing four types of precision bottlenecks and one type of asymmetry that significantly degrade the numerical accuracy of MMAUs (\S\ref{sec:accuracy}). Based on the analysis, we provide software workarounds, mitigation methods, and design suggestions for future MMAUs.

In summary, we make the following contributions:

\begin{itemize}[leftmargin=*]

\item \textbf{A closed-loop feature probing framework.}
We introduce CLFP, a general methodology for deriving bit-accurate models of MMA arithmetic behavior through feature probing and iterative verification and revision.

\item \textbf{Bit-accurate models of GPU MMAUs.}
We construct the first bit-accurate arithmetic behavior models for all floating-point MMA instructions on NVIDIA and AMD GPUs, and implement our models as a GPU-accelerated arithmetic behavior simulator called MMA-Sim.

\item \textbf{MMAU numerical discrepancy analysis.}
We identify the differences in arithmetic behavior across MMAUs and reveal the root causes of numerical discrepancies beyond floating-point non-associativity.

\item \textbf{MMAU numerical accuracy analysis.}
We quantitatively analyze the design choices in MMAUs and their contributions to numerical errors, and provide workarounds and design suggestions.

\item \textbf{Open-source release.}
We open source our implementations and results to facilitate continuous testing, numerical analysis, and design space exploration.

\end{itemize}

\section{Background and Related Work}
\label{sec:background}

\subsection{Numerical Discrepancy and Accuracy Concerns}

Deep neural networks (DNNs) are built on massive numbers of matrix multiplication operations. As models continue to scale, especially in large language model (LLM) workloads, the numerical behavior of the underlying hardware becomes increasingly important. Small floating-point differences can accumulate and amplify across computational layers and iterations, affecting reproducibility, training stability, and model quality \cite{summers_nondeterminism_2021,yuan_understanding_2025}.

Specifically, migrating workloads across hardware vendors or architectural generations can produce different floating-point results even when the same model and input data are used \cite{jezequel_estimation_2015,schlogl_causes_2023,zahid_testing_2024}. Such discrepancies complicate the reproduction of deep learning experiments and create practical challenges in porting workloads across hardware platforms \cite{pham_problems_2020,zhuang_randomness_2022}. Although these discrepancies are often attributed to software factors, such as compiler options, random seeds, mixed-precision policies, and parallel execution orders, they can persist even after software stacks are carefully aligned \cite{laguna_varity_2020,xu_checkpointing_2022,chen_towards_2022}, indicating that conventional software-level explanations are incomplete. To fully diagnose these issues, hardware-specific arithmetic behavior must also be considered.

More importantly, hardware-specific arithmetic behavior can also affect numerical accuracy, with consequences for training stability and model quality. For example, several researchers have reported that insufficient hardware precision in FP8 matrix multiplication degraded training accuracy on NVIDIA Hopper GPUs \cite{deepseek-ai_deepseek-v3_2024,zhao_insights_2025,zhang_sageattention2_2025}. To improve training accuracy, they increased software accumulation precision and suggested increasing hardware accumulation precision to FP32. In addition, PyTorch developers found that DNN training may fail to converge on AMD CDNA2 GPUs because the hardware flushes subnormal values to zero during FP16 matrix multiplication \cite{pytorch_developers_pytorch_2025}. To address this issue, they cast FP16 operands to BF16, thereby increasing the dynamic range at the cost of reduced numerical precision.

\subsection{Black-Box Arithmetic Behavior of MMAUs}

Recent AI accelerators execute deep learning workloads primarily on specialized matrix multiply-accumulate units (MMAUs), such as NVIDIA Tensor Cores \cite{markidis_nvidia_2018} and AMD Matrix Cores \cite{schieffer_rise_2024}. These units provide high throughput and efficiency, but they also make the numerical behavior of DNN computation heavily dependent on hardware-specific matrix multiply-accumulate (MMA) operations. As a result, understanding numerical behavior in modern AI systems increasingly requires understanding the arithmetic behavior implemented by MMAUs.

Although MMAUs share the mathematical definition of MMA operations,
\begin{equation}
\label{eq:mma}
D_{M\times N} = A_{M\times K} \times B_{K\times N} + C_{M\times N},
\end{equation}
their hardware implementations are not necessarily numerically equivalent. The underlying arithmetic behavior of an MMAU is neither standardized nor well documented. Vendors may adopt different implementation details, such as computational orders, accumulation precisions, intermediate rounding modes, fusion granularities, and special-value handling rules, while exposing only MMA instructions or high-level APIs for MMA operations.

Nevertheless, these hidden design choices are critical because they determine how intermediate results are computed and rounded during mixed-precision execution. Consequently, two MMAUs that nominally perform the same MMA operation may still produce different bit-level outputs \cite{fasi_numerical_2021,li_discovery_2024}. Moreover, when numerical discrepancies or accuracy problems arise in software, their root causes are difficult to diagnose because the relevant arithmetic behavior is invisible at the software level.

\subsection{Limitations of Existing Approaches}

Prior work has made important progress in modeling the arithmetic behavior of MMAUs. Feature-probing studies \cite{hickmann_experimental_2019,fasi_numerical_2021,li_fttn_2024,xie_revealing_2025,valpey_smt_2025} have produced approximate models of early-generation Tensor Cores and Matrix Cores through carefully designed numerical tests, and refuted the early model that represented dot products in Tensor Cores as a perfect binary tree of multiplication and addition operations \cite{raihan_modeling_2019}. While these studies provide valuable observations, few studies have derived instruction-level, white-box, bit-accurate models. For this purpose, existing approaches exhibit three important limitations.

First, prior studies inferred arithmetic behavior from MMA APIs rather than directly from instruction-level execution. API-level observations may incorporate software-controlled transformations, and therefore do not distinguish behavior originating in hardware from that introduced by software. For example, Hickmann et al. \cite{hickmann_experimental_2019} attributed the software-dependent reduction order to hardware behavior, and Fasi et al. \cite{fasi_numerical_2021} attributed the hardware-specific FP16 output-rounding behavior to software. 

Second, prior work covered only an incomplete subset of the arithmetic features, which is insufficient for constructing white-box executable models. Consequently, uncharacterized features, such as whether products are normalized before accumulation, which internal summation order is used, and how overflow and underflow are handled, introduce critical ambiguity in bit-accurate modeling.

Third, prior studies did not validate their models at the bit level. As a result, their models may contain false assumptions or inaccurate conclusions. For example, Hickmann et al. \cite{hickmann_experimental_2019} concluded that FP32 subnormal values are treated as zero, which was later disproved by subsequent work \cite{fasi_numerical_2021,li_fttn_2024,valpey_smt_2025}.

Therefore, feature probing alone is insufficient for bit-accurate modeling. The field still lacks a systematic approach for deriving white-box executable models whose semantics are sufficiently complete and whose accuracy is validated against hardware at the bit level.




\section{Methodology and Implementation}
\label{sec:method}

Given a black-box interface of a matrix multiply-accumulate (MMA) operation, we use the closed-loop feature probing (CLFP) framework to build the arithmetic behavior model $\Phi$ such that
\begin{equation}
\label{eq:phi}
    \Phi(A,B,C) = \text{MMA-Interface}(A,B,C),
\end{equation}
for any input matrices $A$, $B$, and $C$ with shapes $M\times K$, $K\times N$, $M\times N$ and data types specified by the interface.

\subsection{CLFP Workflow}

\begin{figure}
    \centering
    \includegraphics[width=1\linewidth]{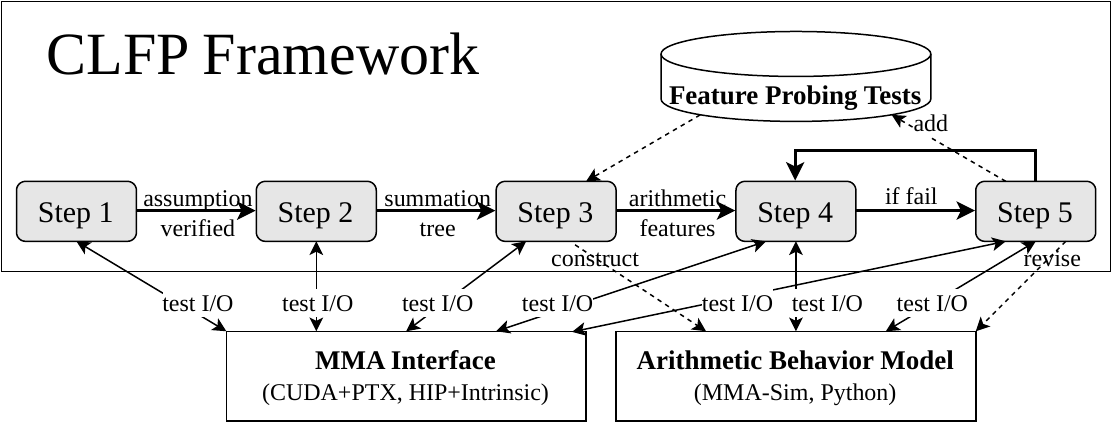}
    \caption{The closed-loop feature probing (CLFP) framework for modeling the arithmetic behavior of the MMA operation. The loop of Steps 4 (verification) and 5 (adding new tests and revising the model) ensures completeness and bit-accuracy.}
    \label{fig:framework}
\end{figure}

As Figure \ref{fig:framework} shows, the workflow of CLFP consists of five major steps. Through arithmetic feature testing and probing, Steps 1 and 2 deterministically decompose the MMA operation into low-level operations, and Step 3 probes the detailed arithmetic behavior of each low-level operation and constructs a model according to the probed behavior. However, the probing may be incomplete, so Step 4 leverages randomized differential testing to verify whether the model composed of the inferred low-level operations matches the output of the MMA interface bit for bit (Equation \ref{eq:phi}). If verification fails, Step 5 transforms the failing case into new feature-probing tests, revises the model accordingly, and returns to Step 4. The loop is repeated until verification succeeds.

\subsubsection{Step 1: Confirming Determinism and Cross-Position Consistency}

Our method assumes that the MMA interface is deterministic. This assumption can be easily verified by randomized testing, and all the MMA interfaces in our experiments have been verified to be deterministic. 

Then, we check whether the computation of
\begin{equation}
\label{eq:dij}
    d_{i,j} = c_{i,j}+\sum_{k=0}^{K-1} a_{i,k}b_{k,j}
\end{equation}
depends on the indices $i$ and $j$. If not, the computation is consistent across output positions, so we can decompose the MMA operation into $M\times N$ independent dot-product-accumulate operations described by Equation \ref{eq:dij} and we only need to model one dot-product-accumulate operation in the remaining steps. Otherwise, we should fall back to per-element modeling.

To verify the computational consistency across output positions, we first generate $2K+1$ random numbers: $a_{0,0}, \dots$, $a_{0,K-1}$, $b_{0,0}, \dots$, $b_{K-1,0}$, and $c_{0,0}$. Then, we let $a_{i,k}=a_{0,k}$, $b_{k,j}=b_{k,0}$, and $c_{i,j}=c_{0,0}$ for all $0\le i<M$, $0\le j<N$, and $0\le k<K$. If the computation does not depend on the indices $i$ and $j$, all $d_{i,j}$ in the output should be bitwise identical; otherwise, elements with different indices may differ in their values.

Through extensive testing, we find that all $d_{i,j}$ in the output matrix are bitwise identical, confirming that each output element in $D$ is computed with consistent arithmetic behavior. Therefore, in the remainder of the paper, we simplify the notation for the dot-product-accumulate operations for any $d_{i,j}$ to
\begin{equation}
\label{eq:d}
d = c + \sum_{k=0}^{K-1}a_{k}b_{k}.     
\end{equation}

\subsubsection{Step 2: Determining Summation Order and Arity}

This step decomposes the dot-product-accumulate operation (Equation \ref{eq:d}) into a combination of summation operations and determines how they are combined.

In particular, Equation \ref{eq:d} can be transformed to 
\begin{equation}
\label{eq:sump}
    d=\sum_{k=0}^K p_k \text{, where }
    p_k=\begin{cases}
        a_kb_k &\text{if $0\le k < K$} \\
        c &\text{if $k=K$}
    \end{cases}.
\end{equation}
The summation order of Equation \ref{eq:sump} can be represented by a summation tree with $K+1$ leaf nodes, where a node with $n$ child nodes represents an $n$-ary summation operation. To infer the tree, we construct the following inputs for Equation \ref{eq:sump}.

\begin{figure}
\centering

\subfloat[Chain of binary summation (e.g., AMD CDNA1 \texttt{mfma\_f32\_16x16x4f32} and NVIDIA \texttt{DMMA.884} instructions).]{%
\begin{minipage}{\linewidth}
\centering
\begin{minipage}{0.45\linewidth}
\centering
\includegraphics[width=\linewidth]{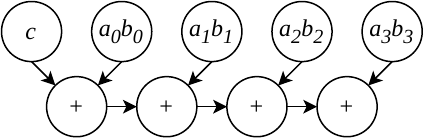}
\end{minipage}
\hfill
\resizebox{0.53\linewidth}{!}{
\centering
\small
\begin{tabular}{ccccc}
\toprule
$U,-U$& $a_1b_1$& $a_2b_2$& $a_3b_3$&$c$\\
\midrule
$a_0b_0$& 2& 1& 0&3\\
$a_1b_1$& & 1& 0&2\\
$a_2b_2$& & & 0&1\\
$a_3b_3$& & & &0\\
\bottomrule
\end{tabular}
}
\end{minipage}
}

\subfloat[Pairwise summation and accumulation (e.g., AMD CDNA2 \texttt{mfma\_f32\_32x32x4bf16\_1k} instruction).]{%
\begin{minipage}{\linewidth}
\centering
\begin{minipage}{0.45\linewidth}
\centering
\includegraphics[width=\linewidth]{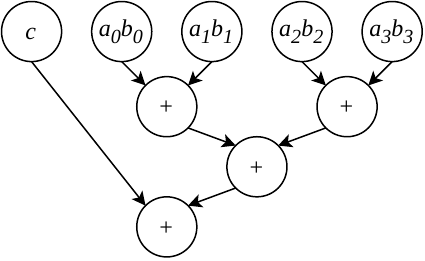}
\end{minipage}
\hfill
\resizebox{0.53\linewidth}{!}{
\centering
\small
\begin{tabular}{ccccc}
\toprule
$U,-U$& $a_1b_1$& $a_2b_2$& $a_3b_3$&$c$\\
\midrule
$a_0b_0$& 3& 1& 1&0\\
$a_1b_1$& & 1& 1&0\\
 $a_2b_2$& & & 3&0\\
$a_3b_3$& & & &0\\
\bottomrule
\end{tabular}
}
\end{minipage}
}

\subfloat[3-term fused summation, high precision (e.g., AMD CDNA1 \texttt{mfma\_f32\_32x32x4bf16} instruction).]{%
\begin{minipage}{\linewidth}
\centering
\begin{minipage}{0.45\linewidth}
\centering
\includegraphics[width=\linewidth]{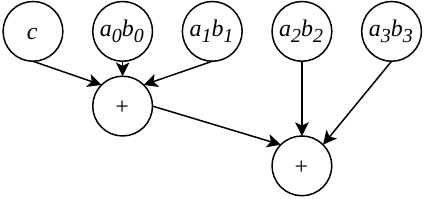}
\end{minipage}
\hfill
\resizebox{0.53\linewidth}{!}{
\centering
\small
\begin{tabular}{ccccc}
\toprule
$U,-U$& $a_1b_1$& $a_2b_2$& $a_3b_3$&$c$\\
\midrule
$a_0b_0$& 3& 1& 1&3\\
$a_1b_1$& & 1& 1&3\\
 $a_2b_2$& & & 3&1\\
$a_3b_3$& & & &1\\
\bottomrule
\end{tabular}
}
\end{minipage}
}

\subfloat[5-term fused summation, high precision (e.g., AMD CDNA3 \texttt{mfma\_f32\_16x16x4\_4b\_bf16} instruction) / limited precision (e.g., NVIDIA \texttt{HMMA.884} instruction).]{%
\begin{minipage}{\linewidth}
\centering
\begin{minipage}{0.45\linewidth}
\centering
\includegraphics[width=\linewidth]{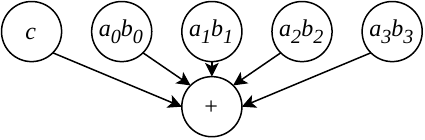}
\end{minipage}
\hfill
\resizebox{0.53\linewidth}{!}{
\centering
\small
\begin{tabular}{ccccc}
\toprule
$U,-U$& $a_1b_1$& $a_2b_2$& $a_3b_3$&$c$\\
\midrule
$a_0b_0$& 3/0& 3/0& 3/0&3/0\\
$a_1b_1$& & 3/0& 3/0&3/0\\
 $a_2b_2$& & & 3/0&3/0\\
$a_3b_3$& & & &3/0\\
\bottomrule
\end{tabular}
}
\end{minipage}
}

\caption{Examples of summation trees (left) and corresponding values of $d^{(i,j)}/v$ (right), where $(i,j)$ denotes the input with $p_i=U$, $p_j=-U$, and all other summands set to $v$.}
\label{fig:tree}
\end{figure}

Adapting the method from FPRev \cite{xie_revealing_2025}, we choose a very large number $U=2^{e_u}$ and a very small number $v=2^{e_v}$ such that 
\begin{equation}
\label{eq:swamp}
    (K-1)v\pm U=\pm U
\end{equation}
in floating-point arithmetic, which means that $\pm U$ can serve as large summands to swamp other small summands in summation. Then, for every $0\le i < j \le K$, we construct the inputs  $a^{(i,j)}_0, \dots, a^{(i,j)}_{K-1}$, $b^{(i,j)}_0, \dots, b^{(i,j)}_{K-1}$, and $c^{(i,j)}$ such that 
\begin{equation}
    p^{(i,j)}_k=\begin{cases}
        U &\text{if $k=i$} \\
        -U &\text{if $k=j$} \\
        v &\text{otherwise}
    \end{cases}.
\end{equation}
The output for these inputs is denoted by $d^{(i,j)}$, and $0 \le d^{(i,j)} < Kv$. The value of $d^{(i,j)}/v$ represents the number of summands not swamped by the large summands $p_i$ and $p_j$, because any summation involving $\pm U$ before the cancellation of $-U+U=0$ results in $\pm U$ (Equation \ref{eq:swamp}) and only the summation after this cancellation contributes to the output, as demonstrated in Figure \ref{fig:tree}. For example, if the summation tree follows Figure \ref{fig:tree}(a) and $i=0$, $j=1$, then the summation is computed as $(((c+a_0b_0)+a_1b_1)+a_2b_2)+a_3b_3=(((v+U)-U)+v)+v=2v$. Therefore, $d^{(i,j)}/v=2$, indicating that only two summands (i.e., $a_2b_2$ and $a_3b_3$) are not swamped by $p_0=a_0b_0$ and $p_1=a_1b_1$.

According to FPRev \cite{xie_revealing_2025}, the values $d^{(i,j)}/v$ for all $0\le i < j \le K$ uniquely determine a summation tree. Therefore, we can infer the summation tree by enumerating all possible trees and checking whether each tree can produce the observed outputs. Alternatively, we can infer the summation tree using a tree realization algorithm based on FPRev. Specifically, FPRev's algorithm assumes that all fused summation operations are performed with limited precision:
\begin{equation}
\label{eq:lp}
\text{FusedSum}_\text{LP}(U,m_1v,\dots,m_{n-2}v,-U) = 0.
\end{equation}
However, we find that high-precision fused summation can break this assumption:
\begin{equation}
\label{eq:hp}
\text{FusedSum}_\text{HP}(U,m_1v,\dots,m_{n-2}v,-U) = \sum_{k=1}^{n-2}m_kv.
\end{equation}
Therefore, FPRev cannot correctly infer the cases based on high-precision fused summation, such as the \texttt{mfma\_f32\_32x32x4bf16} instruction on AMD CDNA1 GPUs. Taking both cases (Equations \ref{eq:lp} and \ref{eq:hp}) into account, we extend the algorithm of FPRev to distinguish between limited-precision and high-precision fused summation by adding corresponding conditional checks. For example, Figure \ref{fig:tree}(d) shows the various values of $d^{(i,j)}/v$ corresponding to the same summation tree with high-precision and limited-precision summation.

\subsubsection{Step 3: Probing Detailed Arithmetic Features}

This step probes the detailed arithmetic behavior of the summation operations in the summation tree constructed in Step 2, aiming at inferring how the operations are performed and building an executable model for the MMA operation. For each arithmetic feature, we construct edge-case inputs and infer the feature from the output. We maintain a collection of feature-probing tests (FPTs). These feature-probing tests are reusable. Once a feature-probing test is manually built, it can be applied to subsequent MMA modeling. The collection is extensive and we showcase the four most important tests here. The remaining tests mainly check binary properties such as whether subnormal values are supported.

\paragraph{FPT 1: Summation Precision}

For every binary addition in the summation tree, we use $\text{Add}(U, \epsilon)$ to test its precision. Specifically, for the addition operation involving $p_i$ and $p_j$, we let $p_i=U$, $p_j=\epsilon=0.5U$, and set other summands to zero. Next, we repeatedly halve $\epsilon$ until the output satisfies $d\neq U+\epsilon$. Then, we conclude that the summation precision of this operation is $e_u-\log_2 \epsilon-1$ fractional bits.

For every $n$-term fused summation, we use $\text{FusedSum}(U, -U, \epsilon)$ to test the precision of the operation. Specifically, for the summation operation involving $p_i$, $p_j$, and $p_k$, we let $p_i=U$, $p_j=-U$, and $p_k=\epsilon=0.5U$, and set other summands to zero. Next, we repeatedly halve $\epsilon$ until $\epsilon=0$ or the output $d\neq \epsilon$. If $d=\epsilon$ remains true until $\epsilon=0$, this operation is exact (as if computed with infinite precision). Otherwise, the summation precision is $e_u-\log_2 \epsilon-1$ fractional bits.

\paragraph{FPT 2: Rounding Mode}

For every summation operation with limited precision, we determine the rounding mode of the operation by testing four sets of inputs: $U+1.5\epsilon$, $U+0.5\epsilon$, $-U-1.5\epsilon$, and $-U-0.5\epsilon$. According to the outputs ($U$ or $U+2\epsilon$, and $-U$ or $-U-2\epsilon$), we can identify the rounding mode as one of ``round toward $+\infty$'' (a.k.a. ``round up'' or RU), ``round toward $-\infty$'' (a.k.a. ``round down'' or RD), ``round toward zero'' (RZ), ``round away from zero'' (RA), and ``round to nearest'' (RN).

If the rounding mode is an RN mode, we test four additional sets of inputs $U+\epsilon$, $U+3\epsilon$, $-U-\epsilon$, and $-U-3\epsilon$. According to the outputs ($U$ or $U+2\epsilon$ or $U+4\epsilon$, and $-U$ or $-U-2\epsilon$ or $-U-4\epsilon$), we can identify the tie-breaking rule as one of ``ties toward $+\infty$'' (a.k.a. ``ties up'' or RNU), ``ties toward $-\infty$'' (a.k.a. ``ties down'' or RND), ``ties toward zero'' (RNZ), ``ties away from zero'' (RNA), ``ties to even'' (RNE), and ``ties to odd'' (RNO).

\paragraph{FPT 3: Internal Summation Order}

A limited-precision $n$-term fused summation operation indicates that intermediate sums are not normalized to floating-point values and only the final sum of $n$ summands is normalized. However, the order of the pre-normalization summation is not specified. To infer the fine-grained summation order within the fused summation operation, we adapt the tests in Step 2 by setting $v=\epsilon$. With this choice, $\pm U$ still swamp individual summands when $\pm U+\epsilon=\pm U$ and the rounding mode applied to $\epsilon$ is RZ, RD, RNE, RNZ, or RND, while the cancellation of $-U+U$ preserves the partial sums produced by earlier additions, yielding $\text{FusedSum}_\text{LP}(U,m_1v,\dots,m_{n-2}v,-U) = \sum_{m_k>1} \text{round}(m_k\epsilon)$. With $d^{(i,j)}/v$ for all $0\le i < j <n$, we can infer the internal summation tree within the operation.

\paragraph{FPT 4: Product Renormalization}

The exponent of a summand determines the position of rounding in a limited-precision $n$-term fused summation operation. For $a_ib_i$, we check whether the product is renormalized before summation by replacing the $\pm U$ with $\pm 1.5\times 1.5\times 2^{e_u}$ in $\text{FusedSum}(U, -U, 2\epsilon)$. If the result remains $2\epsilon$, it implies that the product $\pm 2.25 \times 2^{e_u}$ is not renormalized and the exponent remains $e_u$. Otherwise, if the result changes, it implies that the product is renormalized to $\pm 1.125\times2^{e_u+1}$.

\subsubsection{Step 4: Verification}

The model constructed in Step 3 may need to be revised because the probed arithmetic features may be incomplete and the constructed model may include empirical assumptions that are not verified. Therefore, Step 4 verifies the correctness of the model through randomized differential testing. To extend coverage, our testing includes three types of inputs:
\begin{enumerate}[leftmargin=*]
    \item Random inputs from common distributions, including the normal distribution, the uniform distribution, and typical distributions in DNN workloads such as $N(0,1)+\text{Bernoulli}(0.001)\cdot N(0,10^2)$ \cite{shah_flashattention-3_2024}.
    \item Random inputs with large condition numbers (i.e., $\frac{\sum |p_k|}{|\sum p_k|} \gg 1$) that can trigger catastrophic cancellation \cite{higham_accuracy_2002}. 
    \item Inputs generated from randomized bit-streams, which cover the most diverse inputs including normal numbers with widely varying magnitudes, subnormal numbers, infinities, and NaNs. In practice, this is the most useful input type in our experiments.
\end{enumerate}
Note that edge cases have been integrated as feature-probing tests in Step 3.

If the model is verified to match the output of the MMA interface bit for bit after extensive tests, we complete the modeling for this MMA interface. Otherwise, a failing test case prompts us to revise our model in an additional step.

\subsubsection{Step 5: Generating New Feature-Probing Tests}

When a mismatch occurs in Step 4, we investigate the failing case, transform this case into a new feature-probing test, and revise the model accordingly. With the revised model, CLFP returns to Step 4 to verify the correctness again and repeats this loop until verification passes.

We propose the minimum-failing-subset reduction technique to debug and transform the mismatch case. Specifically, we first zero some input elements and check whether the mismatch persists. After we identify the minimum subset of input elements that induce the mismatch (typically two to four elements), we tune the values of these elements and observe the behavior of both the interface output and the model output. The tuning methods include increasing/decreasing the exponent, truncating/flipping least significant bits, flipping the sign bit, zeroing some fractional bits, etc. The input values at which the outputs switch between matching and mismatching reveal the arithmetic behavior responsible for the mismatch. We convert these inputs to a new feature-probing test that can distinguish the edge cases, and add it to our test collection for future modeling. For example, FPT 4 is generated by the above technique from a mismatch case where the largest $a_ib_i$ satisfies $|\text{Sig}(a_i)\text{Sig}(b_i)|\ge 2$ (where Sig denotes the significand). We revise the model based on the probed behavior as in Step 3, and return to Step 4 for further verification.

\subsection{Method Applicability}

Our CLFP framework is applicable to general MMA operations. Given an interface of an MMA operation, we can build its arithmetic behavior model through CLFP. If the operation interface is at the API level, the derived model reveals the arithmetic behavior of the API. However, without knowing the instruction-level behavior first, it is hard for such a model to distinguish between hardware behavior and software behavior such as programmer intent and compiler transformations that determine which hardware instructions are used and how the instructions are composed.

In comparison, if the interface is at the instruction level, then the derived model reveals the hardware-specific behavior. Therefore, this paper focuses on the hardware matrix multiply-accumulate units (MMAUs) on NVIDIA and AMD GPUs, i.e., Tensor Cores and Matrix Cores. Specifically, this paper applies CLFP to all instruction-level MMA interfaces on ten GPU architectures: NVIDIA Volta (sm70), Turing (sm75), Ampere (sm80), Ada Lovelace (sm89), Hopper (sm90), Blackwell (sm100), and RTX Blackwell (sm120), and AMD CDNA1 (gfx908), CDNA2 (gfx90a), and CDNA3 (gfx942). These architectures represent almost all GPU architectures equipped with Tensor Cores or Matrix Cores as of 2025, which demonstrates the generality of CLFP. 

In future work, we will apply CLFP to more MMAUs on other processors such as CPUs and TPUs \cite{jouppi_-datacenter_2017}, and also experiment with the API-level interfaces.

\subsection{System Implementation}

As Figure \ref{fig:framework} shows, our modeling system consists of three components: the CLFP framework, the MMA interface, and the arithmetic behavior model. All components of our system are open-source on GitHub.

We build the arithmetic behavior models as a software simulator called MMA-Sim. We implement the CLFP framework and MMA-Sim in Python. For NVIDIA devices, we implement the instruction-level MMA interfaces in CUDA and inline assembly (PTX), and verify the correctness of the mappings between PTX instructions (lowest-level programmable assembly) and SASS instructions (hardware instruction set). For AMD devices, we implement the interfaces in HIP and instruction intrinsic functions, and verify the correctness of the function-to-instruction mappings.

We run our experiments on the following GPUs: NVIDIA V100 (Volta), T4 (Turing), A100 (Ampere), RTX 4090 (Ada Lovelace), H100 (Hopper), B200 (Blackwell), and RTX PRO 6000 Blackwell (RTX Blackwell), and AMD MI100 (CDNA1), MI250X (CDNA2), and MI300X (CDNA3).

\subsection{Modeling Efficiency and Simulator Optimization}

CLFP is currently a semi-automated framework. Step 1, Step 2, running feature-probing tests in Step 3, and Step 4 are automated by Python scripts, while Step 5 and model construction in Step 3 are currently manual.

We break down the execution time of CLFP into three parts: (1) GPU execution time, which is less than one millisecond per call, (2) MMA-Sim execution time, which depends on the matrix size of the MMA instruction, and (3) human effort of debugging, adding feature-probing tests, and constructing/revising models.

We note that the execution of MMA-Sim in Step 4 is the most time-consuming part. The original MMA-Sim runs on a single thread and takes approximately 10-20 hours for one million tests in Step 4. As a result, modeling a GPU MMA instruction takes at least one day and sometimes several days depending on the number of unseen arithmetic features.

Therefore, to improve modeling efficiency, we implement GPU acceleration for MMA-Sim. We achieve a speedup of more than $10\times$ and reduce the MMA-Sim execution time from over 10 hours to approximately 1 hour in Step 4.

With the optimized MMA-Sim, modeling a GPU MMA instruction typically takes 2-12 hours, as shown in Table \ref{tab:time}. Steps 1-3 take a small fraction of time for running a fixed number of GPU tests and constructing a model. In Step 4, MMA-Sim takes approximately one hour and GPU execution takes several minutes. If a mismatch occurs, Step 5 is triggered and takes approximately four hours of human effort.

\begin{table}
\centering
\small
\caption{Approximate time of running CLFP with 0-2 revision loops (typical). *MMA-Sim execution time has been reduced from 10-20 hours to 1 hour through optimization.}
\label{tab:time}
\begin{tabular}{ccccc}\toprule
&  \textbf{GPU}&  \textbf{MMA-Sim}&  \textbf{Human}& \textbf{Total}\\\midrule
Step 1&  5 minutes&  0&  0& 5 minutes\\
Step 2&  $<$1 minute&  0&  0& $<$1 minute\\
Step 3&  $<$1 minute&  0&  $<$1 hour& $<$1 hour\\
Step 4&  5 minutes&  1 hour*&  0& 1 hour\\
Step 5&  $<$1 minute&  $<$1 minute&  4 hours& 4 hours\\ 
Total&  10-20 minutes&  1-3 hours&  1-9 hours& 2-12 hours\\ 
\bottomrule
\end{tabular}
\end{table}

The total time of Steps 4 and 5 depends on the number of revision-verification loops. In practice, most instructions can be modeled with zero or one revision. We find that instructions within the same family (e.g., \texttt{HMMA}) exhibit similar arithmetic behavior, and once the first instruction is modeled, existing feature-probing tests are adequate for other instructions of the family and Steps 1-4 run in a single pass. For the first instruction of a new family that has unseen arithmetic features, modeling it typically takes one or two revisions. Only a few instructions require more revisions. The most complex instruction (the FP8 MMA instruction on AMD CDNA3 GPUs) takes ten revisions.

We note that the time for human effort in Step 5 has exceeded MMA-Sim execution time. Therefore, to further accelerate CLFP, we will automate the minimum-failing-subset reduction in Step 5 using Python scripts and try to use AI agents to automate the code writing in model construction and revision.

\section{Derived Bit-Accurate Models}
\label{sec:model}

This section presents our bit-accurate models built through the CLFP framework for all GPU matrix multiply-accumulate (MMA) instructions (also known as matrix fused-multiply-add or MFMA instructions on AMD GPUs). Our models are divided into three categories and eight types based on the elementary operations that compose them, as shown in Table \ref{tab:models}. Each model is composed of \textit{elementary operations}. We define an elementary operation as an $n$-ary floating-point operation $f: \mathbb{F}_1\times \cdots\times \mathbb{F}_n \to \mathbb{F}_\text{out}$ that deterministically maps $n$ floating-point inputs to a single floating-point output. Within an elementary operation, intermediate values are represented in non-$\mathbb{F}_\text{out}$ formats, and intermediate computations are performed using non-$\mathbb{F}_\text{out}$ arithmetic.

\begin{table}
\centering
\small
\caption{Categories of our bit-accurate models for GPU matrix multiply-accumulate units.}
\label{tab:models}
\begin{tabular}{@{}ccc@{}}
\toprule
\textbf{Category} & \textbf{Models} & \textbf{Elementary Ops.} \\ \midrule
MulAdd-based & $\Phi_\text{FTZ-MulAdd}$ & FTZ-Add, FTZ-Mul \\
FMA-based & $\Phi_\text{FMA}$ & FMA \\
FDPA-based & \begin{tabular}[c]{@{}c@{}}$\Phi_\text{E-FDPA}$, $\Phi_\text{T-FDPA}$,\\ $\Phi_\text{ST-FDPA}$, $\Phi_\text{GST-FDPA}$,\\ $\Phi_\text{TR-FDPA}$, $\Phi_\text{GTR-FDPA}$\end{tabular} & \begin{tabular}[c]{@{}c@{}}E-FDPA, T-FDPA,\\ ST-FDPA, GST-FDPA,\\ TR-FDPA, GTR-FDPA\end{tabular} \\ \bottomrule
\end{tabular}
\end{table}

\subsection{Models by Elementary Operations}
\label{sec:ops}

\subsubsection{Model Based on Binary Add and Mul Operations}

The first category of our models is based on binary floating-point multiplication and addition operations. Through CLFP, we find that none of the GPU MMA instructions we examined is composed of standard (IEEE-754 \cite{ieee_ieee_2019}) multiplication and addition operations. In contrast, FP16 and BF16 MMA instructions on AMD CDNA2 GPUs are composed of flush-to-zero addition (FTZ-Add) and flush-to-zero multiplication (FTZ-Mul) operations, which are defined in Algorithm \ref{alg:muladd}. We also find that subnormal MMA inputs are flushed to positive zero, and subnormal outputs of multiplication and addition are flushed to zero with sign preserved. In addition, the order of summation in these instructions is shown in Figure \ref{fig:tree}(b). After the FTZ-Mul operations, each group of $L$ consecutive products ($L\in\{2,4\}$) is summed pairwise using FTZ-Add, and then $c$ and the $K/L$ partial sums are summed sequentially using FTZ-Add. In summary, we build the $\Phi_\text{FTZ-MulAdd}$ model for these instructions as shown in Algorithm \ref{alg:muladd-mma}.

\begin{algorithm}
\small
\caption{FTZ-Add and FTZ-Mul operations.}
\label{alg:muladd}
\begin{algorithmic}[1]
\Require $x,y\in\mathbb{F}$ \Comment{$\mathbb{F}\in\text{\{BF16, FP16, FP32\}}$}
\Ensure $z\in\text{FP32}$
\If{FTZ-Add}
    \State $z\gets \text{RNE}_\text{FP32}(x+y)$ \Comment{round to nearest, ties to even}
\ElsIf{FTZ-Mul} 
    \State $z\gets\text{RNE}_\text{FP32}(x\times y)$
\EndIf
\State $z \gets z \times 0.0$ if $|z|<2^{-126}$ \Comment{flush to zero with sign preserved}
\end{algorithmic}
\end{algorithm}

\begin{algorithm}
\small
\caption{$\Phi_\text{FTZ-MulAdd}$: MMA model based on FTZ-Add and FTZ-Mul.}
\label{alg:muladd-mma}
\begin{algorithmic}[1]
\Require $A\in\mathbb{F}^{M\times K}$, $B\in\mathbb{F}^{K\times N}$, and $C\in\text{FP32}^{M\times N}$ (inputs); $L_\text{max}\in\{2,4\}$ (parameter) \Comment{$\mathbb{F}\in\text{\{BF16, FP16\}}$}
\Ensure $D\in\text{FP32}^{M\times N}$
\State $A\gets \text{FlushSubnormal}(A)$ \Comment{flush input subnormals to $+0.0$}
\State $B\gets \text{FlushSubnormal}(B)$ 
\State $C\gets \text{FlushSubnormal}(C)$ 
\State $L\gets \min(K,L_\text{max})$
\For{\textbf{each} $0\le i<M$ and $0\le j<N$ \textbf{in parallel}}
    \State $d_{i,j}\gets c_{i,j}$
    \For{$k\gets0$ \textbf{to} $K-L$ \textbf{in steps of} $L$}
        \For{\textbf{each} $0\le l<L$ \textbf{in parallel}}
            \State $p_l\gets \text{FTZ-Mul}(a_{i,k+l}, b_{k+l,j})$
        \EndFor
        \State $s \gets \text{FTZ-Add}(p_0, p_1)$
        \If{$L=4$}
            \State $s'\gets\text{FTZ-Add}(p_2, p_3)$
            \State $s\gets \text{FTZ-Add}(s,s')$
        \EndIf
        \State $d_{i,j}\gets \text{FTZ-Add}(d_{i,j}, s)$
    \EndFor
\EndFor
\end{algorithmic}
\end{algorithm}

\subsubsection{Model Based on Ternary FMA Operations}

The second category is based on the ternary floating-point operation fused multiply-add (FMA). Through CLFP, we find that all FP64 MMA instructions on NVIDIA GPUs and all FP64 and FP32 MMA instructions on AMD GPUs are composed of FMA operations. As shown in Algorithm \ref{alg:fma}, this operation takes three floating-point inputs $a$, $b$, and $c$, computes $a\times b+c$ as if with infinite precision, and converts the intermediate result to the output floating-point format using a specific rounding mode $\rho$. The order of operations in these instructions is a chain of FMAs, as shown in Figure \ref{fig:tree}(a). The rounding mode is ``round to nearest, ties to even'' (RNE) by default. On NVIDIA GPUs, it can also be configured as ``round toward zero'' (RZ), ``round up'' (RU), or ``round down'' (RD). In summary, we build the $\Phi_\text{FMA}$ model for these instructions as shown in Algorithm \ref{alg:fma-mma}.

\begin{algorithm}
\small
\caption{FMA operation.}
\label{alg:fma}
\begin{algorithmic}[1]
\Require $a,b,c\in\mathbb{F}$ (inputs); $\rho\in\{\text{RNE}_\mathbb{F}, \text{RZ}_\mathbb{F}, \text{RU}_\mathbb{F}, \text{RD}_\mathbb{F}\}$ (parameter) \Comment{$\mathbb{F}\in\text{\{FP64, FP32\}}$}
\Ensure $d\in\mathbb{F}$
\State $d\gets \rho(a\times b+c)$ \Comment{round to $\mathbb{F}$}
\end{algorithmic}
\end{algorithm}

\begin{algorithm}
\small
\caption{$\Phi_\text{FMA}$: MMA model based on FMA.}
\label{alg:fma-mma}
\begin{algorithmic}[1]
\Require $A\in\mathbb{F}^{M\times K}$, $B\in\mathbb{F}^{K\times N}$, and $C\in\mathbb{F}^{M\times N}$ \Comment{$\mathbb{F}\in\text{\{FP64, FP32\}}$}
\Ensure $D\in\mathbb{F}^{M\times N}$
\For{\textbf{each} $0\le i<M$ and $0\le j<N$ \textbf{in parallel}}
    \State $d_{i,j}\gets c_{i,j}$
    \For{$k\gets 0$ \textbf{to} $K-1$}
        \State $d_{i,j}\gets \text{FMA}(a_{i,k}, b_{k,j}, d_{i,j})$
    \EndFor
\EndFor
\end{algorithmic}
\end{algorithm}

\subsubsection{Models Based on N-ary FDPA Operations}

The third category is based on fused dot-product-add (FDPA) operations. Typically, an FDPA operation takes $2L+1$ floating-point inputs (a pair of vectors of length $L$ and an accumulator) and produces one floating-point output. Through CLFP, we find that most mixed-precision MMA instructions on NVIDIA and AMD GPUs are composed of FDPA operations. Depending on whether $L=K$ or $L<K$, the computation of Equation \ref{eq:d} can be composed of one or more FDPA operations, as shown in Figure \ref{fig:tree}(c) and (d). Therefore, we define the $\Phi_\text{FDPA}$ category in Algorithm \ref{alg:fdpa-mma} for these instructions.

\begin{algorithm}
\small
\caption{$\Phi_\text{FDPA}$: MMA models based on FDPA operations.}
\label{alg:fdpa-mma}
\begin{algorithmic}[1]
\Require $A\in\mathbb{F}_A^{M\times K}$, $B\in\mathbb{F}_B^{K\times N}$, and $C\in\mathbb{F}_C^{M\times N}$ (inputs); $L_\text{max}$ (parameter)
\Ensure $D\in\mathbb{F}_D^{M\times N}$
\State $L\gets \min(K,L_\text{max})$
\For{\textbf{each} $0\le i<M$ and $0\le j<N$ \textbf{in parallel}}
    \State $d_{i,j}\gets c_{i,j}$
    \For{$k\gets0$ \textbf{to} $K-L$ \textbf{in steps of} $L$}
        \State $d_{i,j}\gets \text{FDPA}(a_{i,k}, \dots,a_{i,k+L-1}, b_{k,j}, \dots, b_{k+L-1,j}, d_{i,j})$
    \EndFor
\EndFor
\end{algorithmic}
\end{algorithm}

Through CLFP, we find that different vendors and instructions implement six variants of FDPA operations. Accordingly, we construct six model types in the framework of $\Phi_\text{FDPA}$, each corresponding to a distinct FDPA variant, as described below.

\paragraph{Exact FDPA (E-FDPA)}

We build the exact FDPA (E-FDPA) operation to model BF16 and FP16 MMA instructions on AMD CDNA1 GPUs. Similar to the FMA operation, an E-FDPA operation computes $c+\sum_{k=0}^{L-1}a_kb_k$ as if with infinite precision and converts the intermediate result to the output floating-point format using the RNE rounding mode, as shown in Algorithm \ref{alg:e-fdpa}.

\begin{algorithm}
\small
\caption{E-FDPA operation.}
\label{alg:e-fdpa}
\begin{algorithmic}[1]
\Require $a_0,\dots,a_{L-1},b_0,\dots,b_{L-1}\in\mathbb{F}, c\in\text{FP32}$ \Comment{$\mathbb{F}\in\text{\{BF16, FP16\}}$}
\Ensure $d\in\text{FP32}$
\State $d\gets\text{RNE}_\text{FP32}(c+\sum_{k=0}^{L-1}a_kb_k)$
\end{algorithmic}
\end{algorithm}

\paragraph{Truncated FDPA (T-FDPA)}

We build the truncated FDPA (T-FDPA) operation to model most mixed-precision MMA instructions on NVIDIA GPUs. This operation is parameterized by $L$ (length of the vector), $F$ (precision of the fused summation), and $\rho$ (output rounding mode), and proceeds in three steps as shown in Algorithm \ref{alg:t-fdpa}:
\begin{enumerate}[leftmargin=*]
\item Compute the exact unnormalized products for $L$ pairs of multiplicands, where the signed significands are multiplied using fixed-point arithmetic and the exponents are added using integer arithmetic; 
\item Compute the truncated fused sum for the $L$ products and the accumulator $c$, where these $L+1$ terms are aligned at their largest exponent, their significand bits are truncated to $F$ fractional bits (i.e., RZ to fixed-point format $\mathbb{Q}_F:= 2^{-F}\mathbb{Z}$), and their sum is computed using fixed-point arithmetic;
\item Convert the summation result to the output floating-point format using the rounding mode $\rho$.
\end{enumerate}
We report the probed parameters for each instruction in \S\ref{sec:mapping}.

\begin{algorithm}
\small
\caption{T-FDPA operation.}
\label{alg:t-fdpa}
\begin{algorithmic}[1]
\Require $a_0, \dots, a_{L-1}\in \mathbb{F}_A$, $b_0, \dots, b_{L-1}\in \mathbb{F}_B$, $c\in\mathbb{F}_C$ (inputs); $L$, $F$, $\rho$ (parameters) \Comment{$\mathbb{F}_A,\mathbb{F}_B\in\text{\{TF32, BF16, FP16, FP8, FP6, FP4\}}$}
\Ensure $d\in\mathbb{F}_D$ \Comment{$\mathbb{F}_C,\mathbb{F}_D\in\text{\{FP32, FP16\}}$}
\State \textbf{Step 1: Compute exact products}
\For{\textbf{each} $0\le k <L$ \textbf{in parallel}}
    \State $s_k \gets \text{SignedSig}(a_k) \times \text{SignedSig}(b_k)$ \Comment{exact fixed-point product of signed significands}
    \State $e_k \gets \text{Exp}(a_k)+\text{Exp}(b_k)$ \Comment{sum of exponents}
\EndFor
\State \textbf{Step 2: Compute truncated fused sum of $L+1$ terms}
\State $s_L\gets  \text{SignedSig}(c)$
\State $e_L \gets \text{Exp}(c)$
\State $e_{\text{max}}=\max(e_0, e_1, \dots, e_L)$
\For{\textbf{each} $0\le k \le L$ \textbf{in parallel}}
    \State $s_k' \gets \text{RZ}_{\mathbb{Q}_F}(s_k\times 2^{e_k-e_\text{max}})$ \Comment{align at $e_\text{max}$ and truncate to $F$ fractional bits}
\EndFor
\State $S \gets \sum_{k=0}^L s_k'$ \Comment{exact fixed-point sum}
\State \textbf{Step 3: Convert to floating-point output}
\State $d \gets \rho(S\times 2^{e_\text{max}})$
\end{algorithmic}
\end{algorithm}

\paragraph{Scaled Truncated FDPA (ST-FDPA)}

We build the scaled truncated FDPA (ST-FDPA) operation to model MXFP8/6/4 MMA instructions, as shown in Algorithm \ref{alg:st-fdpa}. ST-FDPA extends T-FDPA by applying per-block scale factors to the inputs before the dot product. It takes $2L+3$ floating-point inputs ($L$ pairs of multiplicands, an accumulator, and two scale factors) and produces one floating-point output. In these instructions, the scale factors are represented in the E8M0 format, and their significands are always  $1.0$. Therefore, based on T-FDPA, we only need to incorporate the exponents of the scale factors in the computation of the products. 

\begin{algorithm}
\small
\caption{ST-FDPA operation.}
\label{alg:st-fdpa}
\begin{algorithmic}[1]
\Require $a_0, \dots, a_{L-1}\in \mathbb{F}_A$, $b_0, \dots, b_{L-1}\in \mathbb{F}_B$, $c\in\text{FP32}$, $\alpha,\beta\in \text{E8M0}$ (inputs); $L$, $F$, $\rho$ (parameters) \Comment{$\mathbb{F}_A,\mathbb{F}_B\in\text{\{FP8, FP6, FP4\}}$}
\Ensure $d\in\text{FP32}$ 
\State \textbf{Step 1: Compute exact products}
\For{\textbf{each} $0\le k <L$ \textbf{in parallel}}
    \State $s_k \gets \text{SignedSig}(a_k) \times \text{SignedSig}(b_k)$ \Comment{exact fixed-point product of signed significands}
    \State $e_k \gets \text{Exp}(a_k)+\text{Exp}(b_k)+\text{Exp}(\alpha)+\text{Exp}(\beta)$ \Comment{sum of exponents}
\EndFor
\State \textbf{Step 2: Compute truncated fused sum of $L+1$ terms}
\State $s_L\gets  \text{SignedSig}(c)$
\State $e_L \gets \text{Exp}(c)$
\State $e_{\text{max}}=\max(e_0, e_1, \dots, e_L)$
\For{\textbf{each} $0\le k \le L$ \textbf{in parallel}}
    \State $s_k' \gets \text{RZ}_{\mathbb{Q}_F}(s_k\times 2^{e_k-e_\text{max}})$ \Comment{align at $e_\text{max}$ and truncate to $F$ fractional bits}
\EndFor
\State $S \gets \sum_{k=0}^L s_k'$ \Comment{exact fixed-point sum}
\State \textbf{Step 3: Convert to floating-point output}
\State $d \gets \rho(S\times 2^{e_\text{max}})$
\end{algorithmic}
\end{algorithm}

\paragraph{Group-Scaled Truncated FDPA (GST-FDPA)}

We build the group-scaled truncated FDPA (GST-FDPA) operation to model specific MXFP4 and NVFP4 MMA instructions. GST-FDPA takes $2L+1+2L/K_\text{block}$ floating-point inputs: a pair of vectors of length $L$, an accumulator, and scale factors for each block of $K_\text{block}$ consecutive elements in each vector. For MXFP4, $K_\text{block}=32$; for NVFP4, $K_\text{block}=16$.

In GST-FDPA, the vector elements are partitioned into groups of size $G$, where $G$ may or may not equal $K_\text{block}$. As shown in Algorithm \ref{alg:gst-fdpa}, GST-FDPA first computes the exact dot product per group in fixed-point arithmetic. Each resulting dot product is then multiplied by the signed significands of the corresponding scale factors, while the corresponding exponent is computed as the sum of the exponents of the scale factors. Finally, GST-FDPA computes the truncated fused sum of the $L/G$ results and $c$, and converts the sum to the output floating-point format.

\begin{algorithm}
\small
\caption{GST-FDPA operation.}
\label{alg:gst-fdpa}
\begin{algorithmic}[1]
\Require $a_0, \dots, a_{L-1}, b_0, \dots, b_{L-1}\in \text{FP4}$, $c\in\text{FP32}$, $\alpha_0, \dots, \alpha_{L/K_\text{block}-1}, \beta_0, \dots, \beta_{L/K_\text{block}-1} \in \mathbb{F}$ (inputs); $L$, $G$, $F$, $\rho$ (parameters) \Comment{$\mathbb{F}\in\{\text{E8M0, UE4M3}\}$}
\Ensure $d\in\text{FP32}$
\State \textbf{Step 1: Compute exact dot products per group}
\For{\textbf{each} $0 \le g < L/G$ \textbf{in parallel}}
    \State $p_g \gets \sum_{k=gG}^{(g+1)G-1} a_{k} \times b_{k}$ \Comment{exact fixed-point dot product}
    \State $s_g \gets p_g \times \text{SignedSig}(\alpha_{\lfloor gG/K_\text{block} \rfloor }) \times \text{SignedSig}(\beta_{\lfloor gG/K_\text{block} \rfloor })$
    \State $e_g \gets \text{Exp}(\alpha_{\lfloor gG/K_\text{block} \rfloor }) + \text{Exp}(\beta_{\lfloor gG/K_\text{block} \rfloor })$ \Comment{sum of exponents}
\EndFor
\State \textbf{Step 2: Compute truncated fused sum of $L/G+1$ terms}
\State $s_{L/G} \gets \text{SignedSig}(c)$
\State $e_{L/G} \gets \text{Exp}(c)$
\State $e_{\text{max}} = \max(e_0, e_1, \dots, e_{L/G})$
\For{\textbf{each} $0 \le g \le L/G$ \textbf{in parallel}}
    \State $s_g' \gets \text{RZ}_{\mathbb{Q}_F}(s_g \times 2^{e_g - e_{\text{max}}})$ \Comment{align at $e_{\text{max}}$ and truncate to $F$ fractional bits}
\EndFor
\State $S \gets \sum_{g=0}^{L/G} s_g'$ \Comment{exact fixed-point sum}
\State \textbf{Step 3: Convert to floating-point output}
\State $d \gets \rho(S \times 2^{e_{\text{max}}})$ 
\end{algorithmic}
\end{algorithm}

\paragraph{Truncated Rounded FDPA (TR-FDPA)}

We build the truncated rounded FDPA (TR-FDPA) operation as shown in Algorithm \ref{alg:tr-fdpa} to model TF32, BF16, and FP16 MMA instructions on AMD CDNA3 GPUs. Compared with T-FDPA, TR-FDPA computes the truncated fused sum for only $L$ products (without $c$). Then, the truncated dot-product result and $c$ are aligned and then rounded down to different precisions ($\mathbb{Q}_{32}$ and $\mathbb{Q}_{F}$), followed by summation and normalization. Specifically, the dot-product result is rounded down to $32$ or more fractional bits. Using more than 32 fractional bits yields the same result because the sum is then normalized to the E8M31 floating-point format using the RD rounding mode. Finally, the result is converted to the output floating-point format.

\begin{algorithm}
\small
\caption{TR-FDPA operation.}
\label{alg:tr-fdpa}
\begin{algorithmic}[1]
\Require $a_0, \dots, a_{L-1}\in \mathbb{F}_A$, $b_0, \dots, b_{L-1}\in \mathbb{F}_B$, $c\in\text{FP32}$ (inputs); $L$, $F$, $\rho$ (parameters) \Comment{$\mathbb{F}_A,\mathbb{F}_B\in\text{\{TF32, BF16, FP16\}}$}
\Ensure $d\in\text{FP32}$
\State \textbf{Step 1: Compute exact products}
\For{\textbf{each} $0\le k <L$ \textbf{in parallel}}
    \State $s_k \gets \text{SignedSig}(a_k) \times \text{SignedSig}(b_k)$ \Comment{exact fixed-point product of signed significands}
    \State $e_k \gets \text{Exp}(a_k)+\text{Exp}(b_k)$ \Comment{sum of exponents}
\EndFor
\State \textbf{Step 2: Compute truncated fused sum of $L$ terms}
\State $e_{\text{dot}}=\max(e_0, e_1, \dots, e_{L-1})$
\For{\textbf{each} $0\le k < L$ \textbf{in parallel}}
    \State $s_k' \gets \text{RZ}_{\mathbb{Q}_F}(s_k\times 2^{e_k-e_\text{dot}})$ \Comment{align at $e_\text{dot}$ and truncate to $F$ fractional bits}
\EndFor
\State $T \gets \sum_{k=0}^{L-1} s_k'$ \Comment{exact fixed-point sum}
\State \textbf{Step 3: Compute rounded sum of two terms}
\State $s_c\gets  \text{SignedSig}(c)$
\State $e_c \gets \text{Exp}(c)$
\State $e_\text{max}\gets \max(e_\text{dot}, e_c)$
\State $T' \gets \text{RD}_{\mathbb{Q}_{32}}(T\times 2^{e_\text{dot}-e_\text{max}})$ \Comment{align at $e_\text{max}$ and round down to $32$ or more fractional bits}
\State $s_c' \gets \text{RD}_{\mathbb{Q}_{F}}(s_c\times 2^{e_c-e_\text{max}})$ \Comment{align at $e_\text{max}$ and round down to $F$ fractional bits}
\State $S\times 2^E\gets \text{RD}_\text{E8M31}(( T'+s_c')\times 2^{e_\text{max}})$ \Comment{normalize and round down to E8M31}
\State \textbf{Step 4: Convert to floating-point output}
\State $d \gets \rho(S\times 2^E)$
\end{algorithmic}
\end{algorithm}

\paragraph{Group-Truncated Rounded FDPA (GTR-FDPA)}

For FP8 MMA instructions on AMD CDNA3 GPUs, we build the group-truncated rounded FDPA (GTR-FDPA) operation. Compared with TR-FDPA, GTR-FDPA computes the truncated fused sums for two groups: the even-indexed products and the odd-indexed products. It then aligns the results of the two groups to their largest exponent, rounds down the aligned significands to $F$ fractional bits, and computes their sum. This dot-product result and $c$ are then summed using a method similar to TR-FDPA's rounded summation with additional truncation, as shown in Algorithm \ref{alg:gtr-fdpa}.

\begin{algorithm}
\small
\caption{GTR-FDPA operation.}
\label{alg:gtr-fdpa}
\begin{algorithmic}[1]
\Require $a_0, \dots, a_{L-1}\in \mathbb{F}_A$, $b_0, \dots, b_{L-1}\in \mathbb{F}_B$, $c\in\text{FP32}$ (inputs); $L$, $F$, $\rho$ (parameters) \Comment{$\mathbb{F}_A,\mathbb{F}_B\in\text{\{FP8\}}$}
\Ensure $d\in\text{FP32}$
\State \textbf{Step 1: Compute exact products}
\For{\textbf{each} $0\le k <L$ \textbf{in parallel}}
    \State $s_k \gets \text{SignedSig}(a_k) \times \text{SignedSig}(b_k)$ \Comment{exact fixed-point product of signed significands}
    \State $e_k \gets \text{Exp}(a_k)+\text{Exp}(b_k)$ \Comment{sum of exponents}
\EndFor
\State \textbf{Step 2: Compute truncated fused sums of $L/2$ terms}
\State $e_{\text{even}}=\max(e_0, e_2, \dots, e_{L-2})$
\State $e_{\text{odd}}=\max(e_1, e_3, \dots, e_{L-1})$
\For{\textbf{each} $0\le k < L/2$ \textbf{in parallel}}
    \State $s_{2k}' \gets \text{RZ}_{\mathbb{Q}_F}(s_{2k}\times 2^{e_{2k}-e_\text{even}})$ \Comment{align at $e_\text{even}$ and truncate to $F$ fractional bits}
    \State $s_{2k+1}' \gets \text{RZ}_{\mathbb{Q}_F}(s_{2k+1}\times 2^{e_{2k+1}-e_\text{odd}})$ \Comment{align at $e_\text{odd}$ and truncate to $F$ fractional bits}
\EndFor
\State $T_\text{even} \gets \sum_{k=0}^{L/2-1} s_{2k}'$ \Comment{exact fixed-point sum}
\State $T_\text{odd} \gets \sum_{k=0}^{L/2-1} s_{2k+1}'$ 
\State \textbf{Step 3: Compute rounded sum of two group sums}
\State $e_\text{dot}=\max(e_\text{even}, e_\text{odd})$
\State $T_\text{even}' \gets \text{RD}_{\mathbb{Q}_{F}}(T_\text{even}\times 2^{e_\text{even}-e_\text{dot}})$ \Comment{align at $e_\text{dot}$ and round down to $F$ fractional bits}
\State $T_\text{odd}' \gets \text{RD}_{\mathbb{Q}_{F}}(T_\text{odd}\times 2^{e_\text{odd}-e_\text{dot}})$ 
\State $T\gets T_\text{even}'+T_\text{odd}'$ 
\State \textbf{Step 4: Compute the final rounded sum}
\State $s_c\gets  \text{SignedSig}(c)$
\State $e_c \gets \text{Exp}(c)$
\State $e_\text{max}\gets \max(e_\text{dot}, e_c)$
\State $T' \gets \text{RD}_{\mathbb{Q}_{32}}(T\times 2^{e_\text{dot}-e_\text{max}})$ \Comment{align at $e_\text{max}$ and round down to $32$ or more fractional bits}
\State $s_c' \gets \text{RD}_{\mathbb{Q}_{F}}(s_c\times 2^{e_c-e_\text{max}})$ \Comment{align at $e_\text{max}$ and round down to $F$ fractional bits}
\If{$e_c<e_\text{max}-F-1$}
    \State $s_c'\gets 0$ \Comment{additional truncation}
\EndIf
\State $S\times 2^E\gets \text{RD}_\text{E8M31}(( T'+s_c')\times 2^{e_\text{max}})$ \Comment{normalize and round down to E8M31 floating-point format}
\State \textbf{Step 5: Convert to floating-point output}
\State $d \gets \rho(S\times 2^E)$
\end{algorithmic}
\end{algorithm}

\subsection{Special Value Handling}

All the elementary operations above handle NaNs and infinities correctly: $\text{NaN}+x=\text{NaN}$, $\text{NaN}\times x=\text{NaN}$, $\pm\infty+y=\pm \infty$, $\pm\infty+\mp\infty=\text{NaN}$, $\pm\infty\times z=\pm\infty\times \text{Sign}(z)$, and $\pm\infty\times 0=\text{NaN}$ for all $x\in\mathbb{F}$, $-\infty<y<\infty$, and $0<|z|<\infty$.

In T-FDPA, ST-FDPA, and GST-FDPA, NaN values are uniformly encoded as 0x7FFFFFFF for FP32 outputs and 0x7FFF for FP16 outputs, and zero-valued $c$ inputs are treated specially as $0.0\times 2^{-108-F}$ for FP32 or $0.0\times 2^{3-F}$ for FP16, rather than the raw representation $0.0\times 2^{-126}$ or $0.0\times 2^{-14}$.

The multiplication in TR-FDPA produces infinity when $|s_k\times 2^{e_k}| \ge 2^{128}$. No other intermediate operation can overflow or underflow. All rounding modes handle overflow or underflow in the standard way.

\subsection{Instruction-to-Model Mapping}
\label{sec:mapping}

For each GPU MMA instruction, we show its corresponding model and parameters derived through CLFP in Table \ref{tab:models-and-params}.

\begin{table*}
\centering
\small
\caption{Models and parameters for all NVIDIA and AMD GPU MMA instructions. In addition, $G=16$ for all instructions modeled by $\Phi_\text{GST-FDPA}$.}
\label{tab:models-and-params}
\begin{tabular}{@{}cccccccc@{}}
\toprule
\textbf{Architecture} & \textbf{Instruction} & \textbf{Model} & \textbf{Input Type} & \textbf{Output Type} & $L_\text{max}$ & $F$ & $\rho$ \\ \midrule
\multirow{2}{*}{Volta} & \multirow{2}{*}{\texttt{HMMA}} & \multirow{2}{*}{$\Phi_\text{T-FDPA}$} & FP16 & FP32 & 4 & 23 & RZ\textsubscript{FP32} \\
 &  &  & FP16 & FP16 & 4 & 23 & RNE\textsubscript{FP16} \\ \midrule
\multirow{2}{*}{Turing} & \multirow{2}{*}{\texttt{HMMA}} & \multirow{2}{*}{$\Phi_\text{T-FDPA}$} & FP16 & FP32 & 8 & 24 & RZ\textsubscript{FP32} \\
 &  &  & FP16 & FP16 & 8 & 24 & RNE\textsubscript{FP16} \\ \midrule
\multirow{4}{*}{Ampere} & \texttt{DMMA} & $\Phi_\text{FMA}$ & FP64 & FP64 & 1 & N/A & Configurable \\ \cmidrule(l){2-8} 
 & \multirow{3}{*}{\texttt{HMMA}} & \multirow{3}{*}{$\Phi_\text{T-FDPA}$} & TF32 & FP32 & 4 & 24 & RZ\textsubscript{FP32} \\
 &  &  & BF16/FP16 & FP32 & 8 & 24 & RZ\textsubscript{FP32} \\
 &  &  & FP16 & FP16 & 8 & 24 & RNE\textsubscript{FP16} \\ \midrule
\multirow{6}{*}{Ada Lovelace} & \texttt{DMMA} & $\Phi_\text{FMA}$ & FP64 & FP64 & 1 & N/A & Configurable \\ \cmidrule(l){2-8} 
 & \multirow{3}{*}{\texttt{HMMA}} & \multirow{3}{*}{$\Phi_\text{T-FDPA}$} & TF32 & FP32 & 4 & 24 & RZ\textsubscript{FP32} \\
 &  &  & BF16/FP16 & FP32 & 8 & 24 & RZ\textsubscript{FP32} \\
 &  &  & FP16 & FP16 & 8 & 24 & RNE\textsubscript{FP16} \\ \cmidrule(l){2-8} 
 & \multirow{2}{*}{\texttt{QMMA}} & \multirow{2}{*}{$\Phi_\text{T-FDPA}$} & FP8 & FP32 & 16 & 13 & RZ\textsubscript{E8M13} \\
 &  &  & FP8 & FP16 & 16 & 13 & RNE\textsubscript{FP16} \\ \midrule
\multirow{6}{*}{Hopper} & \texttt{DMMA} & $\Phi_\text{FMA}$ & FP64 & FP64 & 1 & N/A & Configurable \\ \cmidrule(l){2-8} 
 & \multirow{3}{*}{\texttt{HMMA}/\texttt{HGMMA}} & \multirow{3}{*}{$\Phi_\text{T-FDPA}$} & TF32 & FP32 & 8 & 25 & RZ\textsubscript{FP32} \\
 &  &  & BF16/FP16 & FP32 & 16 & 25 & RZ\textsubscript{FP32} \\
 &  &  & FP16 & FP16 & 16 & 25 & RNE\textsubscript{FP16} \\ \cmidrule(l){2-8} 
 & \multirow{2}{*}{\texttt{QGMMA}} & \multirow{2}{*}{$\Phi_\text{T-FDPA}$} & FP8 & FP32 & 32 & 13 & RZ\textsubscript{E8M13} \\
 &  &  & FP8 & FP16 & 32 & 13 & RNE\textsubscript{FP16} \\ \midrule
\multirow{8}{*}{Blackwell} & \texttt{DMMA} & $\Phi_\text{FMA}$ & FP64 & FP64 & 1 & N/A & Configurable \\ \cmidrule(l){2-8} 
 & \multirow{3}{*}{\texttt{HMMA}/\texttt{UTCHMMA}} & \multirow{3}{*}{$\Phi_\text{T-FDPA}$} & TF32 & FP32 & 8 & 25 & RZ\textsubscript{FP32} \\
 &  &  & BF16/FP16 & FP32 & 16 & 25 & RZ\textsubscript{FP32} \\
 &  &  & FP16 & FP16 & 16 & 25 & RNE\textsubscript{FP16} \\ \cmidrule(l){2-8} 
 & \multirow{3}{*}{\texttt{UTCQMMA}} & \multirow{2}{*}{$\Phi_\text{T-FDPA}$} & FP8/6/4 & FP32 & 32 & 25 & RZ\textsubscript{FP32} \\
 &  &  & FP8/6/4 & FP16 & 32 & 25 & RNE\textsubscript{FP16} \\
 &  & $\Phi_\text{ST-FDPA}$ & MXFP8/6/4 & FP32 & 32 & 25 & RZ\textsubscript{FP32} \\ \cmidrule(l){2-8} 
 & \texttt{UTCOMMA} & $\Phi_\text{GST-FDPA}$ & MXFP4/NVFP4 & FP32 & 64 & 35 & RZ\textsubscript{FP32} \\ \midrule
\multirow{8}{*}{RTX Blackwell} & \texttt{DMMA} & $\Phi_\text{FMA}$ & FP64 & FP64 & 1 & N/A & Configurable \\ \cmidrule(l){2-8} 
 & \multirow{3}{*}{\texttt{HMMA}} & \multirow{3}{*}{$\Phi_\text{T-FDPA}$} & TF32 & FP32 & 8 & 25 & RZ\textsubscript{FP32} \\
 &  &  & BF16/FP16 & FP32 & 16 & 25 & RZ\textsubscript{FP32} \\
 &  &  & FP16 & FP16 & 16 & 25 & RNE\textsubscript{FP16} \\ \cmidrule(l){2-8} 
 & \multirow{2}{*}{\texttt{QMMA}} & \multirow{2}{*}{$\Phi_\text{T-FDPA}$} & FP8/6/4 & FP32 & 32 & 25 & RZ\textsubscript{FP32} \\
 &  &  & FP8/6/4 & FP16 & 32 & 25 & RNE\textsubscript{FP16} \\ \cmidrule(l){2-8} 
 & \texttt{QMMA.SF} & $\Phi_\text{ST-FDPA}$ & MXFP8/6/4 & FP32 & 32 & 25 & RZ\textsubscript{FP32} \\ \cmidrule(l){2-8} 
 & \texttt{OMMA.SF} & $\Phi_\text{GST-FDPA}$ & MXFP4/NVFP4 & FP32 & 64 & 35 & RZ\textsubscript{FP32} \\ \midrule
\multirow{3}{*}{CDNA1} & \multirow{3}{*}{\texttt{v\_mfma}} & $\Phi_\text{FMA}$ & FP32 & FP32 & 1 & N/A & RNE\textsubscript{FP32} \\
 &  & \multirow{2}{*}{$\Phi_\text{E-FDPA}$} & BF16 & FP32 & 2 & N/A & RNE\textsubscript{FP32} \\
 &  &  & FP16 & FP32 & 4 & N/A & RNE\textsubscript{FP32} \\ \midrule
\multirow{5}{*}{CDNA2} & \multirow{4}{*}{\texttt{v\_mfma}} & \multirow{2}{*}{$\Phi_\text{FMA}$} & FP64 & FP64 & 1 & N/A & RNE\textsubscript{FP64} \\
 &  &  & FP32 & FP32 & 1 & N/A & RNE\textsubscript{FP32} \\
 &  & \multirow{2}{*}{$\Phi_\text{FTZ-MulAdd}$} & BF16 & FP32 & 2 & N/A & RNE\textsubscript{FP32} \\
 &  &  & FP16 & FP32 & 4 & N/A & RNE\textsubscript{FP32} \\ \cmidrule(l){2-8} 
 & \texttt{v\_mfma} w/ \texttt{\_1k} suffix & $\Phi_\text{FTZ-MulAdd}$ & BF16 & FP32 & 4 & N/A & RNE\textsubscript{FP32} \\ \midrule
\multirow{5}{*}{CDNA3} & \multirow{5}{*}{\texttt{v\_mfma}} & \multirow{2}{*}{$\Phi_\text{FMA}$} & FP64 & FP64 & 1 & N/A & RNE\textsubscript{FP64} \\
 &  &  & FP32 & FP32 & 1 & N/A & RNE\textsubscript{FP32} \\
 &  & \multirow{2}{*}{$\Phi_\text{TR-FDPA}$} & TF32 & FP32 & 4 & 24 & RNE\textsubscript{FP32} \\
 &  &  & BF16/FP16 & FP32 & 8 & 24 & RNE\textsubscript{FP32} \\
 &  & $\Phi_\text{GTR-FDPA}$ & FP8 & FP32 & 16 & 24 & RNE\textsubscript{FP32} \\ \bottomrule
\end{tabular}
\end{table*}

\subsubsection{NVIDIA Tensor Core Instructions}

For NVIDIA Tensor Cores, the model type depends on the data types of $A$ and $B$ supported by the instruction. FP64 instructions are modeled by $\Phi_\text{FMA}$, MXFP8/6/4 instructions are modeled by $\Phi_\text{ST-FDPA}$, MXFP4/NVFP4 instructions are modeled by $\Phi_\text{GST-FDPA}$, and other instructions with input formats ranging from TF32 to FP4 are modeled by $\Phi_\text{T-FDPA}$.

The specific parameters of the $\Phi_\text{T-FDPA}$, $\Phi_\text{ST-FDPA}$, and $\Phi_\text{GST-FDPA}$ models depend on the architecture, the data types of $A$ and $B$, and the data type of $D$. 

First-generation Tensor Cores on Volta GPUs use a relatively small maximum vector length of $L_\text{max}=4=\frac{8\text{ bytes}}{\text{size of input data type}}$ and a summation precision of $F=23$ fractional bits. The output rounding mode $\rho$ depends on the data type of $D$: RZ for FP32 and RNE for FP16.

Turing, Ampere, and Ada Lovelace GPUs double $L_\text{max}$ to $\frac{16\text{ bytes}}{\text{size of input data type}}$ and increase $F$ to 24. However, Ada Lovelace introduces FP8 formats with $F=13$ and $\rho\in\{\text{RZ}_\text{E8M13},\text{RNE}_\text{FP16}\}$, reflecting the reduced summation precision for FP8.

Hopper GPUs extend this trend, doubling $L_\text{max}$ to $\frac{32\text{ bytes}}{\text{size of input data type}}$, increasing $F$ to 25 for non-FP8 inputs, and keeping $F=13$ for FP8 inputs.

The most recent Blackwell and RTX Blackwell GPUs add FP6, FP4, and MXFP variants while maintaining similar rounding conventions and the same $L_\text{max}$ and $F$ values as Hopper GPUs for most formats. For FP8 formats, $F$ increases to $25$, and $\rho$ changes from $\text{RZ}_\text{E8M13}$ to $\text{RZ}_\text{FP32}$ for FP32 outputs. For MXFP4 or NVFP4 inputs, GST-FDPA operations specifically support a larger $F=35$, indicating a dedicated configuration for extremely low-precision formats.

\subsubsection{AMD Matrix Core Instructions}

For AMD Matrix Cores, the model type depends not only on the input data types of $A$ and $B$, but also on the architecture, with the exception of FP64 and FP32 instructions, which can be consistently modeled by $\Phi_\text{FMA}$ across architectures.

On CDNA1 GPUs, BF16 and FP16 instructions can be modeled by $\Phi_\text{E-FDPA}$ with different maximum vector lengths $L_\text{max}$ depending on the input data type. On CDNA2 GPUs, BF16 and FP16 instructions shift to $\Phi_\text{FTZ-MulAdd}$ with various $L_\text{max}$ parameters depending on the input type and the instruction suffix. CDNA3 GPUs introduce other FDPA variants for lower-precision formats, including $\Phi_\text{TR-FDPA}$ for TF32, BF16, and FP16, and $\Phi_\text{GTR-FDPA}$ for FP8. Their parameters follow a consistent pattern across input types: $L_\text{max}=\frac{16\text{ bytes}}{\text{size of input data type}}$, $F=24$, and $\rho=\text{RNE}_\text{FP32}$. The $L_\text{max}$ and $F$ parameters are identical to those of NVIDIA Ampere GPUs.

\section{Discrepancy Analysis}
\label{sec:discrepancy}

Our white-box models provide a hardware perspective on numerical discrepancies. Across vendors and architectures, only FP64 and FP32 MMA instructions can maintain consistent numerical behavior because they are composed of the same FMA operations performed in the same sequential order. In contrast, mixed-precision MMA instructions use different elementary operations with different parameters, resulting in pervasive numerical discrepancies.

Based on our models, we demonstrate how the discrepancies arise using the following example input:
\begin{equation}
\label{eq:ABC}
\begin{split}
&A= ( \mathbf{a}, \mathbf{0}, \dots, \mathbf{0})^T, 
B= ( \mathbf{b}, \mathbf{0}, \dots, \mathbf{0}), C=(\mathbf{c}, \mathbf{0}, \dots, \mathbf{0}),\\
&\mathbf{a}=(-2^{13},  -0.5, -0.25, -0.125, 0, \dots, 0)^T, \\
&\mathbf{b}=(2^{10},  1, 1, 1, 0, \dots, 0)^T, \mathbf{c}=(2^{23}, 0, \dots, 0)^T. \\
\end{split}
\end{equation}
The output $d_{0,0}$ varies across architectures and instructions, as shown in Table \ref{tab:d00}. The value of $d_{0,0}$ is computed as the sum of $c=2^{23}$, $a_0b_0=-2^{23}$, $a_1b_1=-0.5$, $a_2b_2=-0.25$, and $a_3b_3=-0.125$. For FP64 and FP32 instructions, all devices produce the exact result $-0.875$ following the sequential FMA computation. 

\begin{table}
\centering
\small
\caption{The divergent results of different MMA instructions for the same input in Equation \ref{eq:ABC}. In addition, all FP64/FP32 instructions produce $d_{0,0}=-0.875$.}
\label{tab:d00}
\begin{tabular}{cccc}\toprule
\textbf{Architecture}&\textbf{TF32/BF16 Instr.}&  \textbf{FP16 Instr.}& \textbf{FP8 Instr.}\\\midrule
Volta&N/A&  $0.0$& N/A\\
Turing&N/A&  $-0.5$& N/A\\
Ampere&$-0.5$&  $-0.5$& N/A\\
Ada Lovelace&$-0.5$&  $-0.5$& $0.0$\\
Hopper&$-0.75$&  $-0.75$& $0.0$\\
Blackwell&$-0.75$&  $-0.75$& $-0.75$\\
RTX Blackwell&$-0.75$&  $-0.75$& $-0.75$\\
CDNA1&$-0.875$&  $-0.875$& N/A\\
CDNA2&$-0.375$ or $0.0$&  $0.0$& N/A\\
CDNA3&$-0.5$& $-0.5$&$-1.0$\\ \bottomrule
\end{tabular}
\end{table}

On Volta GPUs, the truncated summation with $F=23$ (the number of fractional bits in the accumulator) can only produce $0.0$ because $-0.5$ and the summands of smaller magnitude are truncated to zero. The same behavior occurs for FP8 (E5M2) instructions on Ada Lovelace and Hopper GPUs, where $F=13$. On Turing, Ampere, and Ada Lovelace GPUs, the truncated summation with $F=24$ produces $-0.5$ because $-0.25$ and $-0.125$ are truncated to zero. On Hopper, Blackwell, and RTX Blackwell GPUs, the truncated summation with $F=25$ produces $-0.75$ because only $-0.125$ is truncated to zero.

Non-truncated summation on CDNA1 GPUs produces the exact result $-0.875$. On CDNA2 GPUs, the FP16 instructions and the BF16 instructions with the ``\texttt{\_1k}'' suffix use the summation order 
$2^{23} + ((-2^{23}-0.5)+(-0.25-0.125))
=2^{23} + (-2^{23}-0.375)
=2^{23} - 2^{23}$
and produce $0.0$. The BF16 instructions without the ``\texttt{\_1k}'' suffix use the summation order 
$(2^{23} + (-2^{23}-0.5))+(-0.25-0.125)
=(2^{23} - 2^{23})-0.375
=0 - 0.375$
and produce $-0.375$. On CDNA3 GPUs, the TF32, BF16, and FP16 instructions compute the fused truncated sum of $-2^{23}$, $-0.5$, $-0.25$, and $-0.125$ with $F=24$, producing the intermediate result $-2^{23}-0.5$, and then add it to $2^{23}$, resulting in $-0.5$. The FP8 (E5M2) instructions on CDNA3 GPUs perform fused summation with $F=24$ in two groups, $-2^{23}-0.25$ and $-0.5-0.125$, producing $-2^{23}$ and $-0.625$. They then compute the rounded sum of these two partial sums with $F=24$ and produce $-2^{23}-1$ because $-0.625$ is rounded down to $-1$. Finally, they add the rounded sum $-2^{23}-1$ to $c=2^{23}$, yielding $-1$.

\section{Accuracy Analysis}
\label{sec:accuracy}

Leveraging our white-box models, we analyze the numerical accuracy of MMA instructions by quantifying every source of numerical error. Based on the analysis, we identify design choices that affect numerical precision and accuracy. These design choices explain the root causes of known hardware-related numerical issues on NVIDIA Hopper and AMD CDNA2 GPUs and expose potential numerical issues on other GPU architectures.

\subsection{Sources of Numerical Errors}

All sources of numerical errors in MMA instructions are summarized in Table \ref{tab:error}. The errors accumulate additively throughout the computation.

\begin{table}
\centering
\small
\caption{Sources and upper bounds of numerical error.}
\label{tab:error}
\begin{tabular}{ccc}
\toprule
\textbf{Elementary Op.}& \textbf{Error Source}& \textbf{Error Bound}\\
\midrule
\multirow{3}{*}{FTZ-Add/Mul} & Input FTZ & $2^{-126}$ or $2^{-14}$ \\
& Add/Mul & $0.5\text{ ulp}$\\
& Output FTZ & $2^{-126}$ \\
\midrule
FMA, E-FDPA& Output rounding & $0.5\text{ ulp}$ or $1\text{ ulp}$ \\
\midrule
\multirow{2}{*}{T-FDPA, others} & Fused summation & $(L+1)2^{e_\text{max}-F}$ \\
& Output rounding & $0.5\text{ ulp}$ or $1\text{ ulp}$ \\ \bottomrule
\end{tabular}
\end{table}

In $\Phi_\text{FTZ-MulAdd}$, the error bound for subnormal flushing corresponds to the smallest normal number of the floating-point format: $2^{-14}$ for FP16 and $2^{-126}$ for FP32 or BF16. In the FTZ-Add and FTZ-Mul operations, the error bound for addition and multiplication is $0.5\text{ ulp}$ (unit in the last place \cite{higham_accuracy_2002}) of the FP32 result, equal to $0.5\times 2^{e_\text{result}-23}$.

In $\Phi_\text{FMA}$ and $\Phi_\text{E-FDPA}$, since the internal computation of FMA and E-FDPA semantically operates at infinite precision, only output rounding contributes an error with an upper bound of $0.5\text{ ulp}$ (the RNE rounding mode) or $1\text{ ulp}$ (other rounding modes).

In $\Phi_\text{T-FDPA}$ and other $\Phi_\text{FDPA}$ variants, the errors come from two sources: fused summation and output rounding. In fused summation, the $L+1$ summands are eventually aligned at $e_\text{max}$ and rounded to $F$ fractional bits. This introduces an error of at most $(L+1)2^{e_\text{max}-F}$. In output rounding, the error bound depends on $\rho$: RNE\textsubscript{FP32} and RNE\textsubscript{FP16} have an error bound of $0.5\text{ ulp}$, while RZ\textsubscript{FP32} and RZ\textsubscript{E8M13} have an error bound of $1\text{ ulp}$.

\subsection{Precision and Accuracy Concerns}

Most of the error sources we identified have error bounds comparable to those of standard FP32 operations, i.e., $0.5\text{ ulp}_\text{FP32}=0.5\times  2^{e_\text{result}-23}$. However, a few exhibit substantially larger error bounds and therefore constitute ``precision bottlenecks''. In addition, we find that certain instructions on AMD CDNA3 GPUs violate the odd symmetry $\Phi(-A,B,-C)=-\Phi(A,B,C)$. Such asymmetry can also degrade numerical accuracy. We summarize these concerns in Table \ref{tab:issues}.  

\begin{table}
\centering
\small
\caption{Concerns regarding numerical precision (top) and accuracy (bottom).}
\label{tab:issues}
\begin{tabular}{cc}
\toprule
\textbf{Architectures and Instructions}&\textbf{Concerns}\\
\midrule
AMD CDNA2, FP16-input&Input FTZ\\
NVIDIA Ada Lovelace and Hopper, FP8-input&Small $F$\\
NVIDIA Ada Lovelace and Hopper, FP8-input&$\rho=\text{RZ}_\text{E8M13}$\\
All NVIDIA architectures, FP16-output&$\rho=\text{RNE}_\text{FP16}$\\
\midrule
AMD CDNA3, TF32/BF16/FP16/FP8-input&Asymmetry\\ \bottomrule
\end{tabular}
\end{table}

\subsubsection{Input FTZ of FP16 Subnormals}
\label{sec:issue-ftz}

Although both input FTZ and output FTZ exist in $\Phi_\text{FTZ-MulAdd}$, the output FTZ poses lower risk because the output type is FP32 and the error is at most $2^{-126}$. In contrast, the input FTZ may introduce a significantly larger error (at most $2^{-14}$) for FP16 input. This explains why training stability of FP16 deep neural networks may degrade on AMD CDNA2 GPUs \cite{pytorch_developers_pytorch_2025}. 

\subsubsection{Reduced Precision in FP8 MMA}
\label{sec:issue-precision}

According to our models, the \texttt{QMMA} instructions on NVIDIA Ada Lovelace GPUs and the \texttt{QGMMA} instructions on NVIDIA Hopper GPUs employ only $F=13$ fractional bits for fused summation within the T-FDPA operation and use $\rho=\text{RZ}_\text{E8M13}$ when the output type is FP32. By contrast, their \texttt{HMMA} or \texttt{HGMMA} instructions use $F=24$ (Ada Lovelace) or $F=25$ (Hopper), similar to the precision of FP32 (E8M23), and use $\rho=\text{RZ}_\text{FP32}$ when the output type is FP32. Consequently, the error bound of the truncated fused summation in \texttt{QMMA}/\texttt{QGMMA} is higher than that in \texttt{HMMA}/\texttt{HGMMA} by a factor of $2^{24-13}=2048$ or $2^{25-13}=4096$, and the error bound of the output rounding is higher by a factor of $2^{23-13}=1024$. These account for the degraded training accuracy of FP8 large language models on Hopper GPUs \cite{deepseek-ai_deepseek-v3_2024,zhao_insights_2025}.

\subsubsection{Limited Output Precision}
\label{sec:issue-output}

As shown in Table \ref{tab:models-and-params}, some \texttt{HMMA}, \texttt{HGMMA}, \texttt{UTCHMMA}, \texttt{QMMA}, and \texttt{QGMMA} instructions on NVIDIA GPUs produce FP16 outputs with $\rho=\text{RNE}_\text{FP16}$, whose error bound is $0.5\text{ ulp}_\text{FP16}=0.5\times 2^{e_\text{result}-10}$. Therefore, even when the internal fused summation has FP32-like precision, the overall precision is bounded by the output format.

\subsubsection{Asymmetric Rounding}
\label{sec:issue-asymmetry}

We note that most MMA instructions use symmetric rounding modes such as RZ and RNE internally. However, we find that the mixed-precision \texttt{v\_mfma} instructions on AMD CDNA3 GPUs use RD, an asymmetric rounding mode, for internal summation, which may introduce numerical bias.

\begin{figure}[ht]
\centering
\includegraphics[width=0.75\linewidth]{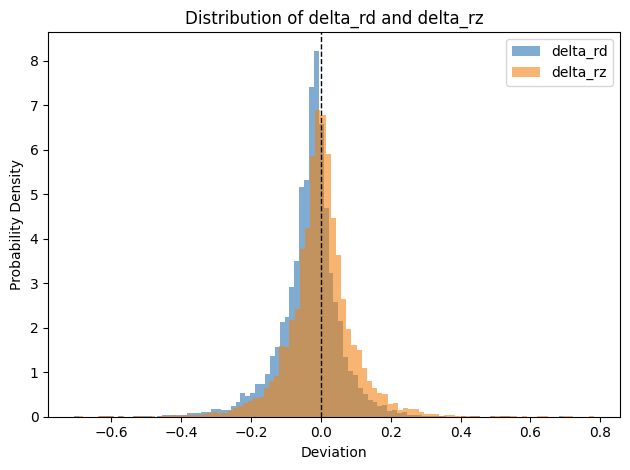}
\caption{Distributions of $\delta_\text{RD}$, the numerical deviation of the AMD CDNA3 FP16 MMA instruction, which uses the ``round down'' (RD) mode internally, and $\delta_\text{RZ}$, the numerical deviation of a hypothetical variant that replaces the internal RD operations with ``round toward zero'' (RZ) operations.}
\label{fig:delta}
\end{figure}

To demonstrate the numerical bias, we simulate the CDNA3 \texttt{v\-\_mfma\-\_f32\-\_32x32x8\-\_f16} instruction with $\Phi_\text{TR-FDPA}$ and a hypothetical variant \texttt{v\-\_mfma\-\_f32\-\_32x32x8\-\_f16\-\_rz} using a modified model $\Phi_\text{TR-FDPA}'$ that replaces the RD mode in the internal fused summation with RZ. We sample $C$ from the standard normal distribution $N(0,1)$ and sample $A$ and $B$ from the normal distribution $1000\times N(0, 1)$. Then, we obtain the CDNA3 output $D_\text{RD}=\Phi_\text{TR-FDPA}(A,B,C)$ and the output of the RZ variant $D_\text{RZ}=\Phi_\text{TR-FDPA}'(A,B,C)$. We also compute the accurate FP64 result $D_\text{FP64}=A\times B+C$, and measure the deviations $\delta_\text{RD}=D_\text{RD}-D_\text{FP64}$ and $\delta_\text{RZ}=D_\text{RZ}-D_\text{FP64}$. As Figure \ref{fig:delta} shows, the distribution of $\delta_\text{RD}$ exhibits a negative bias while the distribution of $\delta_\text{RZ}$ is symmetric.

\subsection{Workarounds}
\label{sec:workaround}

\paragraph{Workaround 1: Scaling Inputs}
We note that the CDNA2 FP16 FTZ issue (\S\ref{sec:issue-ftz}) manifests prominently for input elements of small magnitude. Therefore, we suggest scaling up the inputs to prevent FP16 subnormals and scaling down the output to restore the original scale. For example, consider multiplying $A_{128\times 128}$ sampled from $N(0, 0.001^2)$ with $B_{128\times 128}$ sampled from $N(0, 1)$. This workaround reduces the mean squared error (MSE) relative to the FP64 result from $7.632\times 10^{-9}$ to $1.591\times 10^{-18}$. Since the computational precision of CDNA2 FP16 MMA is similar to FP32 except for the input FTZ (Table \ref{tab:error}), this workaround preserves the high computational precision. Compared with PyTorch's workaround \cite{pytorch_developers_pytorch_2025}, which casts FP16 to BF16 and introduces precision loss during conversion, our workaround has better numerical precision.

\paragraph{Workaround 2: MMA without $C$}
We note that the numerical bias on AMD CDNA3 (\S\ref{sec:issue-asymmetry}) occurs when $A$ and $B$ contain large-magnitude numbers and the elements of $C$ are negative numbers with relatively small magnitudes. Therefore, developers can use Matrix Cores only for matrix multiplication by setting $C=0$, and use the general compute units for full FP32-precision matrix accumulation to mitigate this issue. Using $128\times 128\times 128$ matrix multiplication as an example, this workaround reduces the mean deviation relative to the FP64 result from $-1.352\times 10^{-6}$ to $+7.467\times 10^{-8}$ for FP16 inputs sampled from the normal distribution $N(0, 10^2)$.

\paragraph{Workaround 3: Changing Instructions}
On NVIDIA GPUs, developers should avoid the instructions with precision bottlenecks (\S\ref{sec:issue-precision}, \S\ref{sec:issue-output}) and use higher-precision alternative instructions instead. FP16-output MMA instructions should be replaced by FP32-output instructions. For the $128\times 128\times 128$ matrix multiplication example above, this can reduce the MSE from $0.247$ to $4.481\times 10^{-8}$ (or $0.0551$ if FP32 results are converted back to FP16). FP8 matrix multiplication on NVIDIA Ada Lovelace or Hopper GPUs should be implemented with FP16 MMA instructions instead of FP8 instructions when accuracy takes precedence over throughput. This can reduce the MSE of the example above from $0.0103$ to $5.569\times 10^{-11}$ for FP8 inputs.

\subsection{Suggestions for Future MMAU Design}

First, MMAU designers should carefully scrutinize the numerical accuracy of hardware designs. According to our modeling results, despite diverse arithmetic operations and configuration choices across architectures, most instructions have FP32-like numerical accuracy. To achieve this, hardware designers should avoid the designs in Table \ref{tab:issues} and maintain FP32-level precision across all computation steps. Second, as symmetric rounding modes are preferable, we also suggest adopting the RZ mode because RZ can be implemented as truncation for sign-magnitude or ones' complement encoding, which is far cheaper than other rounding modes.

\section{Conclusion}
\label{sec:conclusion}

This paper introduces the first bit-accurate models of the arithmetic behavior of matrix multiply-accumulate units (MMAUs) on NVIDIA and AMD GPUs through our closed-loop feature probing (CLFP) approach. With the white-box models, we analyze the root causes of MMAU numerical discrepancies and numerical errors, providing insights into software mitigation and future hardware designs. This work is open-source, enabling continuous testing, transparent numerical analysis, and accuracy-aware design space exploration.

\appendix
\section{Artifact Appendix}

\subsection{Abstract}

Our artifacts consist of 
(1) the closed-loop feature probing (CLFP) framework and the feature-probing test (FPT) collection used to derive bit-accurate arithmetic models for GPU matrix multiply-accumulate (MMA) instructions, 
(2) MMA-Sim, a Python software simulator that implements the derived models (\S\ref{sec:model}) and reproduces the bit-accurate output of every tested MMA instruction, and 
(3) the CUDA/PTX and HIP instruction-level MMA interfaces for ten NVIDIA and AMD GPU architectures.
These artifacts provide fine-grained implementation details of our methods and algorithms that are not exhaustively covered in the paper, and enable independent verification and reproduction of our results.
We open-source the artifacts at \url{https://github.com/microsoft/MMA-Sim}, and provide a Jupyter notebook for easy reproduction of our results.

\subsection{Artifact check-list (meta-information)}

{\small
\begin{itemize}
  \item {\bf Binary: } Python packages \texttt{mmasim} and \texttt{mmasim-kernels} are available on PyPI.
  \item {\bf Run-time environment: } Jupyter notebook.
  \item {\bf Hardware: } A GPU with Tensor Cores or Matrix Cores.
  \item {\bf Metrics: } Bit-exact match between MMA-Sim output and hardware output; execution time; numerical accuracy.
  \item {\bf How much disk space is required (approximately): } Less than 10 GB.
  \item {\bf How much time is needed to prepare the workflow (approximately): } Less than 1 hour.
  \item {\bf How much time is needed to complete the experiments (approximately): } About 1 hour.
  \item {\bf Publicly available: } Yes.
  \item {\bf Code licenses (if publicly available): } MIT license.
\end{itemize}
}

\subsection{Description}

\subsubsection{How to access}

\begin{itemize}
  \item Source code: \url{https://github.com/microsoft/MMA-Sim}.
  \item Jupyter notebook: \url{https://github.com/peichenxie/MMA-Sim-MICRO2026-AE}.
\end{itemize}

\subsubsection{Hardware dependencies}

Software-only experiments with MMA-Sim can run on CPUs without a GPU.
Running hardware-specific experiments requires a GPU with Tensor Cores or Matrix Cores.
For example, the Jupyter notebook requires a Turing-architecture GPU such as NVIDIA T4.

\subsubsection{Software dependencies}

We recommend running the Jupyter notebook on online platforms like Colab.
Alternatively, installing our artifact locally requires (1) a recent Linux distribution, (2) compatible GPU drivers and CUDA/ROCm toolkits, and (3) PyTorch 2.7+ with CUDA or ROCm support.

\subsection{Installation}

Our notebook includes installation commands.
For local experiments, run \texttt{pip install mmasim mmasim-kernels}.

\subsection{Evaluation and expected results}

We demonstrate the following three experiments in both our notebook and source code.

\paragraph{Verification of bit-accuracy.} To confirm that MMA-Sim matches hardware output bit for bit, run the \path{tests/equivalence/test_<your_target_architecture>.py} script. The script is expected to pass all tests without any errors.

\paragraph{Evaluation of CLFP execution time.} To measure the execution time of CLFP for one MMA instruction, run the \path{tests/probe/clfp_example_step<1-4>.py} scripts with a timer. All tests are expected to pass without any errors, and the execution time is expected to be consistent with that reported in Table \ref{tab:time}.

\paragraph{Reproduction of discrepancy and error analysis.} To reproduce the discrepancy example of \S\ref{sec:discrepancy}, run the \path{examples/discrepancy/discrepancy_example.py} script. The output is expected to be consistent with Table \ref{tab:d00}. To evaluate the numerical errors for the workaround methods in \S\ref{sec:workaround}, run the scripts in \path{examples/accuracy/}. The measured errors are expected to be close to the values reported in \S\ref{sec:workaround}.

\bibliographystyle{IEEEtran}
\bibliography{references}

\end{document}